\documentclass[aps,twocolumn,prd,showpacs,showkeys,preprintnumbers,superscriptaddress,nobibnotes,floatfix,longbibliography,notitlepage,nofootinbib]{revtex4-1}
\pdfoutput=1
\usepackage{amsmath}
\usepackage{amsfonts}
\usepackage{amssymb}
\usepackage{mathrsfs}
\usepackage{color}
\usepackage{slashed}
\usepackage{graphicx}
\usepackage{xcolor}
\usepackage{url}
\usepackage[colorlinks=true,allcolors=blue]{hyperref}
\usepackage{multirow}
\usepackage{upgreek}
\usepackage[capitalise]{cleveref}
\crefformat{footnote}{#2\footnotemark[#1]#3}
\usepackage{braket}
\usepackage[normalem]{ulem}
\usepackage{fontawesome}

\usepackage{siunitx}

\DeclareSIUnit \s {\second}
\DeclareSIUnit \ns {\nano\second}
\DeclareSIUnit \mus {\micro\second}
\DeclareSIUnit \ms {\milli\second}
\DeclareSIUnit \MB {\mega\byte}
\DeclareSIUnit \GB {\giga\byte}
\DeclareSIUnit \TB {\tera\byte}
\DeclareSIUnit \PB {\peta\byte}
\DeclareSIUnit \Mbps {\mega\bit/\s}
\DeclareSIUnit \Gbps {\giga\bit/\s}
\DeclareSIUnit \Tbps {\tera\bit/\s}
\DeclareSIUnit \Pbps {\peta\bit/\s}
\DeclareSIUnit \kton {\kilo\tonne} 
\DeclareSIUnit \kt {\kilo\tonne}
\DeclareSIUnit \kty {\kilo\tonne-\year}
\DeclareSIUnit \Mt {\mega\tonne}
\DeclareSIUnit \eV {\electronvolt}
\DeclareSIUnit \keV {\kilo\electronvolt}
\DeclareSIUnit \MeV {\mega\electronvolt}
\DeclareSIUnit \GeV {\giga\electronvolt}
\DeclareSIUnit \TeV {\tera\electronvolt}
\DeclareSIUnit \PeV {\peta\electronvolt}
\DeclareSIUnit \EeV {\exa\electronvolt}
\DeclareSIUnit \m {\meter}
\DeclareSIUnit \cm {\centi\meter}
\DeclareSIUnit \nm {\nano\meter}
\DeclareSIUnit \in {\inchcommand}
\DeclareSIUnit \km {\kilo\meter}
\DeclareSIUnit \kV {\kilo\volt}
\DeclareSIUnit \kW {\kilo\watt}
\DeclareSIUnit \MW {\mega\watt}
\DeclareSIUnit \MHz {\mega\hertz}
\DeclareSIUnit \mrad {\milli\radian}
\DeclareSIUnit \year {years}
\DeclareSIUnit \POT {POT}
\DeclareSIUnit \sig {$\sigma$}
\DeclareSIUnit\parsec{pc}
\DeclareSIUnit\lightyear{ly}
\DeclareSIUnit\foot{ft}
\DeclareSIUnit\ft{ft}
\DeclareSIUnit \ppb{ppb}
\DeclareSIUnit \ppt{ppt}
\DeclareSIUnit \samples{S}
\DeclareSIUnit \pe{PE}
\DeclareSIUnit \GeVmwe{GeV/mwe}
\DeclareSIUnit \mwe{mwe}

\newcommand{\enu}{\E_\enu}

\usepackage[bottom]{footmisc}

\newcommand{\TB}[1]{{\color{teal}#1}}

\newcommand{\ie}{{\it i.e.}}

\newcommand{\eg}{{\it e.g.}}

\newcommand{\Losc}{L_{\text{osc}}}
\newcommand{\Lcoh}{L_{\text{coh}}}

\begin{document}

\title{Impact of Wave Packet Separation in Low-Energy Sterile Neutrino Searches}

\author{Carlos A.~Arg{\"u}elles}
\email{carguelles@fas.harvard.edu}
\affiliation{Department of Physics \& Laboratory for Particle Physics and Cosmology, Harvard University, Cambridge, MA 02138, USA}

\author{Toni Bert\'olez-Mart\'inez}
\email{antoni.bertolez@fqa.ub.edu}
\affiliation{Departament de F\'isica Qu\`antica i Astrof\'isica and Institut de Ci\`encies del Cosmos, Universitat de Barcelona, Diagonal 647, E-08028 Barcelona, Spain}

\author{Jordi Salvado}
\email{jsalvado@icc.ub.edu}
\affiliation{Departament de F\'isica Qu\`antica i Astrof\'isica and Institut de Ci\`encies del Cosmos, Universitat de Barcelona, Diagonal 647, E-08028 Barcelona, Spain}

\date{\today}

\begin{abstract}
Light sterile neutrinos have been motivated by anomalies observed in short-baseline neutrino experiments.
Among them, radioactive-source and reactor experiments have provided evidence and constraints, respectively, for electron neutrino disappearance compatible with an eV-scale neutrino.
The results from these observations are seemingly in conflict. 
This paper brings into focus the assumption that the neutrino wave packet can be approximated as a plane wave, which is adopted in all analyses of such experiments. 
We demonstrate that the damping of oscillation due to decoherence effects, \eg, a finite wave packet size, solves the tension between these electron-flavor observations and constraints.
\end{abstract}

\maketitle
\section{Introduction}
The observation of an excess of electron antineutrino events in the Liquid Scintillator Neutrino Detector (LSND)~\cite{LSND:1996ubh,LSND:2001aii} in the mid-1990s started a broad experimental program to confirm this signal.
The simplest explanation of the excess is that it is due to the presence of a fourth neutrino, whose flavor state does not participate in the Standard Model weak interactions, and whose mass splitting is on the order of $\SI{1}\eV^2$.
Given this as the explanation of the LSND observation, we then expect that correlated signals should be present at different baselines and energies but at a similar ratio of baseline-to-energy of approximately $\SI{1}\GeV/\si\km$.

\begin{figure}[ht]
    \centering
    \includegraphics[width = \linewidth]{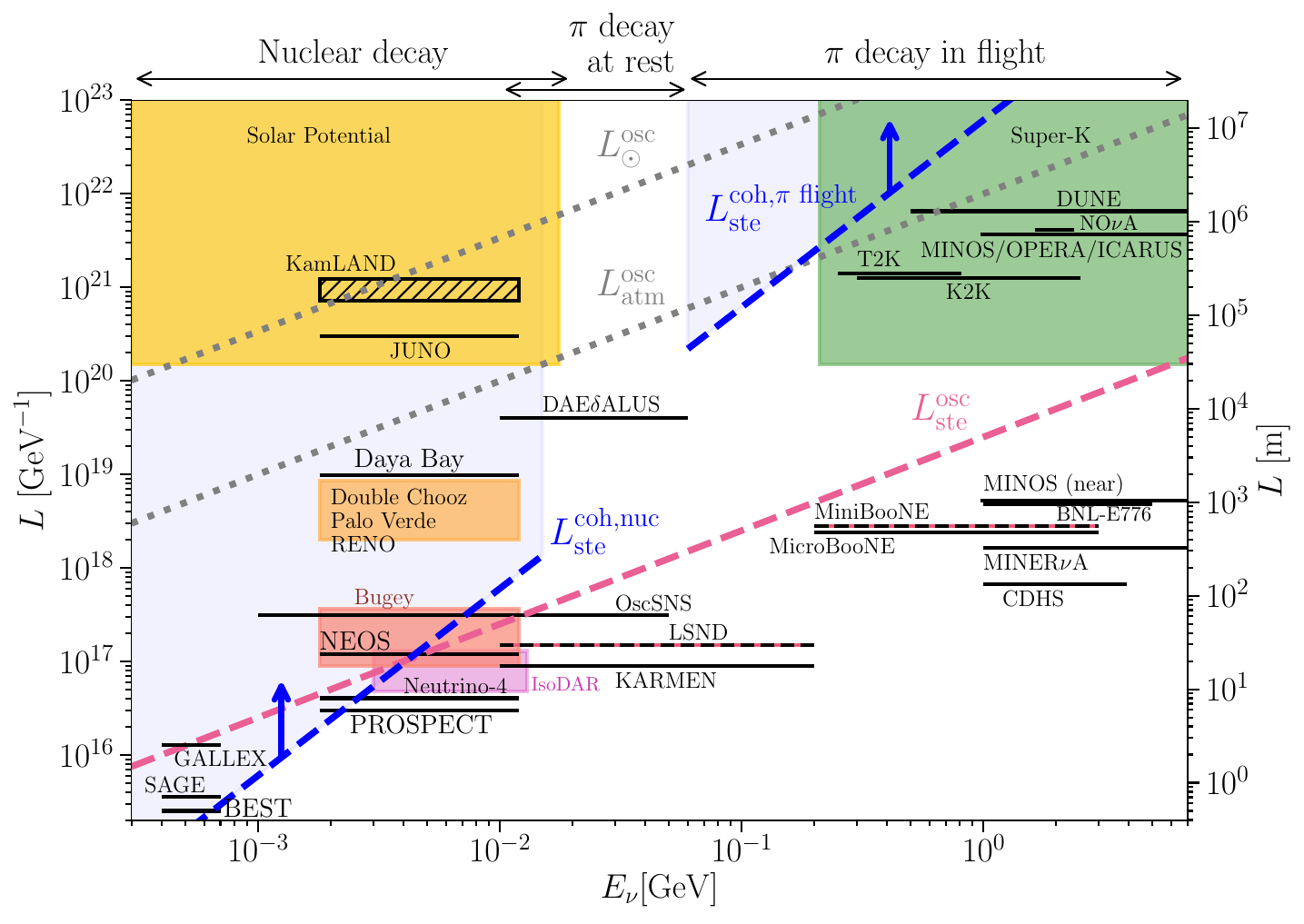}
    \caption{\textbf{\textit{Overview of the solar potential, neutrino experiments, and relevant scales.}} $L^{\text{osc}}$ (dotted gray and dashed pink) and $L^{\text{coh}}$ (dashed blue) are computed from~\cref{eq:characteristic_lengths} using $\Delta m_{41}^2 = \SI{1}\eV^2$ and $\sigma_x = 2.1\times 10^{-4}\text{ nm}$ for $L^{\text{coh,nuc}}_{\text{ste}}$, and $\sigma_x = 10^{-11}$ m for $L^{\text{coh,}\pi\text{ flight}}_{\text{ste}}$~\cite{Jones:2014sfa}.
    Decoherence effects are expected at $L\gtrsim L^{\text{coh}}$.
    Matter effects important for solar oscillations in the Sun are under the region so-called solar potential.}
    \label{fig:all-experiments}
\end{figure}

Experiments searching for these signatures have been performed with energies ranging from MeV to TeV and baselines from a few meters to the diameter of the Earth, as shown in~\cref{fig:all-experiments}.
These experiments use neutrinos produced predominantly by three means: nuclear decay in the MeV range, pion decay at rest at the $\SI{100}\MeV$ scale, and pion or kaon decay in flight in the highest energy range.
In the lowest energy range, gallium experiments study the production rate of inverse beta decay on $^{71}$Ga from an intense electron neutrino source~\cite{Bahcall:1994bq,Bahcall1997,SAGE:1998fvr,Barinov:2021asz, Giunti:2010zu,Abdurashitov:2005tb}. Also at MeV energies, reactor experiments have performed searches for the presence of electron antineutrino disappearance by comparing observations to theoretical predictions of the rates~\cite{Mention:2011rk} or by searching for oscillatory patterns in measurements performed at different positions~\cite{DayaBay:2016qvc, DayaBay:2016ggj,NEOS:2016wee,Stereo66,solid,neutrino42021,PhysRevLett.125.191801, Andriamirado2021,Barinov:2021mjj,DANSS:2018fnn,Danilov:2021oop}.
All these low-energy experiments have yielded confirmatory signals that range in significance from $\sim 2$ to more than 5 sigma but at the same time have yielded constraints that contradict these observations, specially when taking into account solar neutrino analysis~\cite{Berryman:2021yan, Goldhagen:2021kxe}.
In the intermediate energy range, the MiniBooNE~\cite{MiniBooNE:2013uba,MiniBooNE:2018esg} experiment has reported the appearance of electron-neutrino-like events compatible with the LSND observation at a significance of 4.8 sigma.
Operating in the same beam, recently the MicroBooNE collaboration has published measurements of electron neutrino events under various interaction channels~\cite{MicroBooNE:2021jwr,MicroBooNE:2021nxr,MicroBooNE:2021rmx,MicroBooNE:2021sne}.
When this data is interpreted in the context of a light sterile neutrino, weak signals for electron-neutrino disappearance are observed~\cite{Denton:2021czb} and weak constraints on the MiniBooNE region are obtained~\cite{Arguelles:2021meu,MiniBooNE:2022emn}.
Finally, in the highest energy range, the MINOS+ collaboration has placed very strong constraints on muon-neutrino disappearance, while the IceCube Neutrino Observatory observes a mild signal~\cite{MINOS:2017cae,IceCube:2016rnb,IceCube:2020tka,IceCube:2020phf}.
This is a very confusing situation that, when studied in the context of global fits, results in the conclusion that the inconsistencies between the datasets rule out the light sterile neutrino interpretation of LSND~\cite{Dentler:2018sju,Giunti:2019aiy,Diaz:2019fwt,Boser:2019rta}.

In this paper, we point out that the above-mentioned conclusion, specifically about the apparent contradiction between reactor experiments and radioactive sources, has overlooked an important fact that could resolve the tension.
When deriving the results quoted above, the experiments assume that the neutrino state is a plane wave.
It is well-known that the plane-wave (PW) theory of neutrino oscillations~\cite{Eliezer:1975ja,Fritzsch:1975rz,Bilenky:1976yj} is a simplified framework that upon careful inspection contains apparent paradoxes~\cite{Akhmedov:2019iyt,Giunti:2003ax,Akhmedov:2009rb}.
These can be resolved by introducing the wave packet (WP) formalism~\cite{Nussinov:1976uw,Kayser:1981ye,Kiers:1995zj,Beuthe:2001rc,Akhmedov:2012uu,Akhmedov:2017mcc, Bahcall:1994bq,Giunti:1997wq}.
The applicability of the plane-wave approximation has been studied in detail for the standard mass-squared differences~\cite{Akhmedov:2009rb,Beuthe:2001rc,Giunti:2007ry,Bernardini:2004sw,Naumov:2010um} and has been shown to be a good approximation for current and future neutrino experiments.
However, this has not been shown to be the case for mass-squared differences relevant to the LSND observation~\cite{LSND:1996ubh}.
The correctness of the PW approximation depends on the wave packet width, which varies with the neutrino production and detection processes. For example, in the case of pion decay in flight the wave packet size has been quantitatively estimated~\cite{Jones:2014sfa}, and as such it is inconsequential to the light sterile neutrino analyses.
This is seen in~\cref{fig:all-experiments}, where we compare the oscillation length and the coherence length.
In the case of pion decay at rest or production from nuclear reactors or radioactive sources, this has not been precisely calculated. 
In particular, for nuclear reactors, it has been suggested that the relevant scales for the neutrino wave packet width could be~\cite{DayaBay:2016ouy}: the typical size of the beta-decaying nuclei ($\sim 10^{-5}\si\nm$), the interatomic spacing that characterizes the fuel ($\sim 0.01 - \SI{1}\nm$ for uranium
), or the inverse of the antineutrino energy ($\sim 10^{-4}\si\nm$), or the mean free path of the parent nucleus ($\sim 10^{2}\si\nm$)~\cite{Akhmedov:2022bjs}. Most of these values are not definitive quantitative results~\cite{Jones:2022cvh}. As a matter of fact, a recent study following the formalism of open quantum systems states that the wave packet width should lie in the $0.01-0.4\si\nm$ range~\cite{Jones:2022hme}.

Taking an agnostic viewpoint, our current knowledge is limited to bounds from experiments measuring the standard oscillation scales, which set it to be no smaller than $2.1\times  10^{-4}\si\nm$~\cite{deGouvea:2021uvg,DayaBay:2016ouy}. 

In this work, we focus on the low-energy region, where searches using electron antineutrinos from nuclear reactors and radioactive sources are performed~\cite{DayaBay:2016qvc, DayaBay:2016ggj,NEOS:2016wee,PhysRevLett.125.191801, Andriamirado2021,Barinov:2021asz}.
We will show how the plane wave approximation breaks for values of the wave packet size currently allowed~\cite{deGouvea:2021uvg} and how introducing this formalism produces observable effects. 
It is worth mentioning that the damping of oscillations in neutrino physics is not exotic but an expected phenomenon in some scenarios.
On the one hand, a precise enough measurement of the kinematics of the final states in the production region may effectively measure the mass of the outgoing neutrino. This effect is referred to as quantum damping and is believed to be very small for the production of neutrinos~\cite{Stodolsky:1998tc}.
On the other hand, the wave packets may separate during propagation due to their different masses.
This effect of decoherence is strictly equivalent to taking into account the proper energy uncertainty in the production and detection processes~\cite{Akhmedov:2022bjs}.
The latter is related to the spatial and time localization of the interaction or, equivalently, the uncertainty in the measurement of the neutrino energy.
Since both these phenomena are physically indistinguishable, in this work we consider the energy resolution claimed by experiments and add a decoherence effect that introduces a damping of the oscillations. This addition can either be understood as a separation of the wave packets or as an underestimation of the energy uncertainties claimed by experiments.
Finally, the macroscopical production and detection regions averaging also produces the same effect but is already considered in experimental analyses.

Caveats or fundamental physics unknowns in the wave packet size estimations or any exotic physics can enlarge the damping effect.
For this reason, we choose the smallest wave packet size allowed by present bounds obtained from studies of standard oscillations in nuclear reactor experiments~\cite{deGouvea:2021uvg,DayaBay:2016ouy}.
Notice then that this must be robust under the most exotic scenario since it involves the same production and detection process and does not rely on any assumptions.
Moreover, the chosen value is preferred by experiments at 90\%~C.L.~\cite{deGouvea:2021uvg}. 

The allowed size of the wave packet, together with the larger sterile mass value, brings us to the main points of this paper.
First: experimental results may need to consider the decoherence effects arising from the WP formalism, which might produce damped oscillations.
Second: these effects may modify both signals from radioactive sources and exclusion regions from nuclear reactors and can indeed alleviate part of the tension between them.

The remainder of this paper is organized in the following sections: \textit{Formalism}, where we introduce the wave packet formalism for neutrino oscillations;
\textit{Impact on neutrino experiments}, where we show the impact of the finite wave packet size from sterile neutrino searches by the Daya Bay, NEOS, BEST, and PROSPECT experiments and we discuss the results; and,
finally, in \textit{Conclusions}, where we summarize our main findings.

\section{Formalism} In the plane wave formalism, a propagating neutrino is modeled with perfectly defined momentum.
However, this approximation cannot fully convey the physics effects we mention earlier.
Here we are going to parametrize the damping of the oscillations by a length scale $\sigma_x$ that is usually referred to as the wave packet size~\cite{Giunti:1991ca,Giunti:1997wq,Nussinov:1976uw,Kayser:1981ye,Kiers:1995zj,Beuthe:2001rc,Akhmedov:2012uu,Akhmedov:2017mcc,Akhmedov:2017xxm}. 

The length $\sigma_x$ can appear explicitly if we assume the wave packets to be Gaussian. This allows for more concise analytical results, such as the oscillation probability
\begin{align} \label{eq:general_prob}
P_{\alpha\beta} =& \sum_{i=1}^n |U_{\alpha i}|^2|U_{\beta i}|^2 + 2\text{Re} \sum_{j>i}U_{\alpha i}U_{\alpha j}^*U_{\beta i}^*U_{\beta j}\times \\ \nonumber 
\times& \exp\left\{-2\pi i\frac{L}{\Losc^{ij}}-2\pi^2\left(\frac{\sigma_x}{\Losc^{ij}}\right)^2 - \left(\frac{L}{\Lcoh^{ij}}\right)^2\right\},
\end{align} where $U_{\alpha i}$ are the neutrino mixing matrix elements and $L$ the experiment baseline.
Here we have defined
\begin{equation} \label{eq:characteristic_lengths}
    \Losc^{ij} = \frac{4\pi E}{\Delta m^2_{ji}}\quad \text{and}\quad
    \Lcoh^{ij} = \frac{4\sqrt{2}E^2\sigma_x}{\Delta m^2_{ji}},
\end{equation}the oscillation and coherence lengths, respectively.
Note that~\cref{eq:general_prob} is the usual oscillation probability, with two additional terms in the exponential, which dampen the oscillation and only appear if we follow the WP formalism.

The term $({\sigma_x}/{\Losc^{ij}})^2$ inside the exponential in~\cref{eq:general_prob} is significant when $\sigma_x \sim \Losc^{ij}$.
In this regime, the wave packet width from production and/or detection is so large that it does not allow distinguishing between mass eigenstates. This results in washed-out oscillations.
Most experiments, such as the ones studied here, work in the limit $\sigma_x \ll \Losc^{ij}$, such that this term is negligible. 
Therefore, we will ignore it from here on. 

On the other hand, the term $(L/\Lcoh^{ij})^2$ is significant when $L\gtrsim \Lcoh^{ij}$.
This term can be understood as the decoherence arising from the separation of the mass eigenstates during their propagation at different velocities.
The larger $L$, the more separation, the more decoherence and the more dampening of the oscillations.
As explained before, this term may be absorbed in the response function of the detector and thus could also be interpreted as a worsening of its energy resolution~\cite{Akhmedov:2022bjs}.
Note from~\eqref{eq:characteristic_lengths} that the dampening increases with smaller $\sigma_x$ and larger $\Delta m^2_{ji}$.
Thus, this effect may be important when studying mass-squared differences relevant to the LSND observation, since they are typically orders of magnitude larger than the standard ones.

\begin{figure}[ht]
    \centering
    \includegraphics[width = \linewidth]{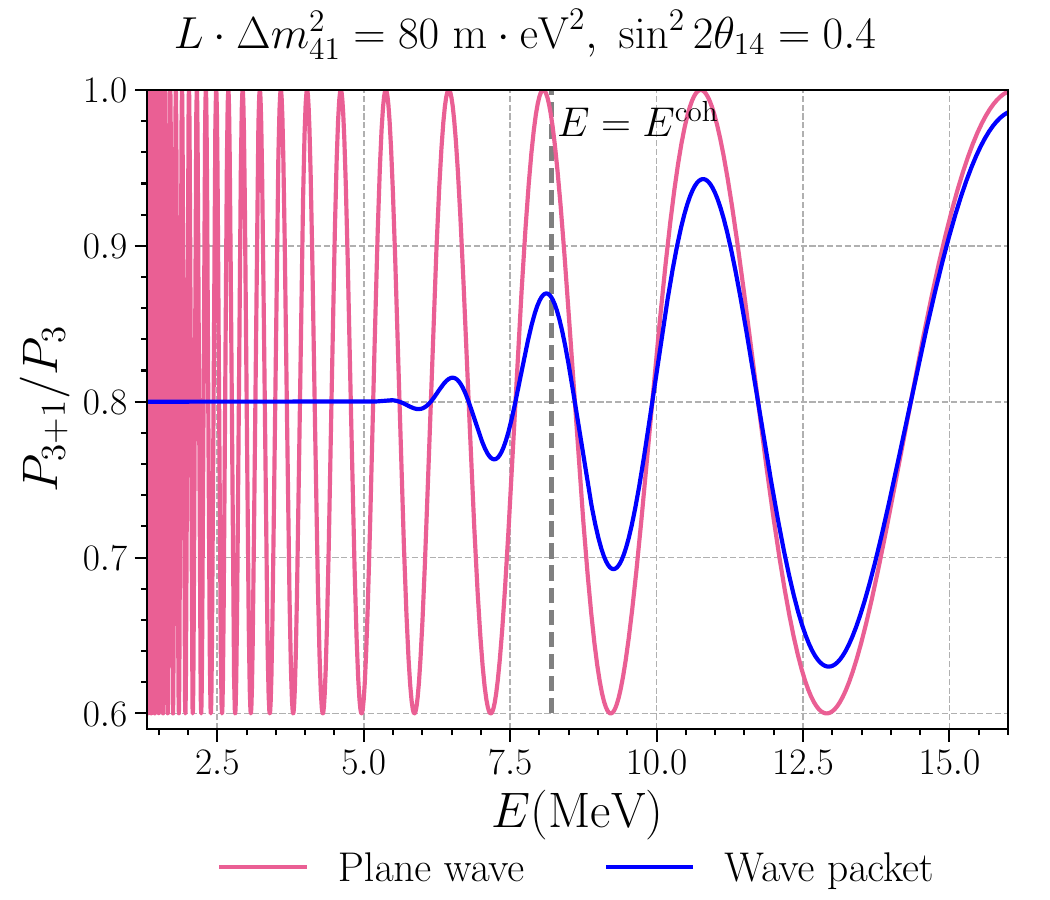}
    \caption{\textbf{\textit{Illustration of the wave packet effect.}} Plot of the oscillation probability ratio for $\sigma_x = 2.1\times 10^{-4}\si\nm$.
    The $y$ axis represents the ratio between the 3+1 and the 3 neutrino oscillation probabilities, in the PW formalism (pink) and in the WP one (blue).
    The effect demonstrated here would appear for $\mathcal{O}(0.1 \text{ eV}^2)$ sterile at the Daya Bay baselines, or $\mathcal{O}(1 \text{ eV}^2)$ sterile at the NEOS or PROSPECT baselines.
    The energy where $L^{\text{coh}}_{\text{ste}} = L^{\text{osc}}_{\text{ste}}$ is defined as $E^{\text{coh}}_{\text{ste}}$ and is independent of the sterile neutrino mass.
    This energy is indicated as a vertical dashed line.}
    \label{fig:probability}
\end{figure}

Nuclear decay experiments study the electron antineutrino survival probability, $P(\nu_e\to\nu_e) \equiv P_{ee}$.
Following from~\cref{eq:general_prob}, and considering here only sterile and atmospheric oscillations for concision, this is given by\footnote[2]{In our analysis, we consider the entire expression.} 
\begin{equation} \label{eq:surv_prob}
    P_{ee} \approx 1- \sin^22\theta_{14}\Delta_{41} - \sin^22\theta_{13}\Delta_{31},
\end{equation}
where we have defined
\begin{equation}\label{eq:Delta}
    \Delta_{ji} = \frac{1}{2}\left(1-\cos\frac{L\Delta m_{ji}^2}{2E}\exp\left\{-\frac{L^2(\Delta m^2_{ji})^2}{32E^4\sigma_x^2}\right\} \right).
\end{equation}
While in the PW limit, $\Delta_{ji} = \sin^2(L\Delta m_{ji}^2/4E)$.
Then, this is the analogous result to Ref.~\cite{DayaBay:2016qvc}, but taking into account decoherence effects.
The difference between both results with and without decoherence effects are shown in~\cref{fig:probability} for parameters motivated by the LSND observation and wave packet at the current constraints.

For illustration purposes, we show with a vertical line the energy $E_{\text{coh}} = \sqrt{L\Delta m_{ji}^2/(4\sqrt{2}\sigma_x)}$ for which the exponential argument of the coherence suppression term is equal to 1.
Three different regimes can be clearly distinguished.
At low energies, oscillations are very fast and cannot be resolved given the experimental energy resolution, resulting in averaging of the oscillations that cannot be distinguished from the decoherence effect.
At energies close to $E^{\text{coh}}$, decoherence can produce an observable effect that in principle can be measured and distinguished from other oscillation features.
Finally, at high energies, the decoherence effect becomes less important and eventually is a small correction to the oscillation amplitude.

In order to understand the potential impact of the decoherence effect, it is useful to compare the different relevant scales.
\cref{fig:all-experiments} shows several oscillation experiments compared to the sterile oscillation scale ($L_{\text{ste}}^{\text{osc}}$) and the decoherence scale ($L_{\text{ste}}^{\text{coh}}$); in both cases the parameters corresponds to the $\Delta m_{41}^2 = \SI{1}\eV^2$ and the current best constrain for $\sigma_x=2.1\times10^{-4}{\rm nm}$~\cite{deGouvea:2021uvg}.
For experiments with baselines smaller than $L_{\text{ste}}^{\text{coh}}$, decoherence can be neglected, while experiments with large baselines will experience complete decoherence.
Notice that the effect of not resolving fast oscillations experimentally is from an observational point of view identical to a decoherence effect, meaning that an experiment far above the $L_{\text{ste}}^{\text{osc}}$ line would also be effectively decoherent, and no effect due to $L_{\text{ste}}^{\text{coh}}$ would be manifest.
This narrows the region of interest for the decoherence of light sterile neutrinos to the low-energy region and in particular to the reactor and radioactive sources experiments. 

\section{Impact on neutrino experiments}
To show the impact of the wave packet separation we choose the smallest value allowed for the wave packet size, $\sigma_x=2.1\times 10^{-4}{\rm nm}$~\cite{deGouvea:2021uvg}, and perform analyses searching for sterile neutrinos with and without the plane wave approximation.
In this work, we use this bound both in reactor and gallium experiments for simplicity, even though they need not have the same wave packet size.
In our global analysis, we consider the null results from Daya Bay~\cite{DayaBay:2016ggj,DayaBay:2016qvc}, NEOS~\cite{NEOS:2016wee}, and PROSPECT~\cite{Andriamirado2021} and the anomalous results observed from radioactive sources by BEST~\cite{Barinov:2021asz}.
This is not an exhaustive list of affected experiments, but they are sufficient to cover the regions of interest illustrated in~\cref{fig:all-experiments}.
The aim of this paper is not to perform a global fit in the WP formalism, but to illustrate its phenomenology in low-energy sterile searches in the context of decoherence effects.

The Daya Bay experiment data has been fit using a test statistic ($\mathcal{TS}^{\rm DayaBay}(\theta_{14},\Delta m^2_{41},\vec \alpha)$) based on a Poisson log-likelihood with nuisance parameters that account for the flux systematic uncertainties ($\vec\alpha$).
The number of expected events has been computed following~\cite{Dentler:2017tkw}, assuming an electron antineutrino flux from Huber and Mueller~\cite{Huber:2011wv,Mueller:2011nm}.
However, these fluxes come with associated uncertainties, and cannot reproduce the observations~\cite{Dentler:2017tkw,Huber2016} with complete accuracy.
To minimize the dependence on the flux model, we introduce a nuisance parameter for each energy bin, which must be the same for the three experimental halls of the Daya Bay experiment.

\begin{figure}[ht]
    \centering
    \includegraphics[width = \linewidth]{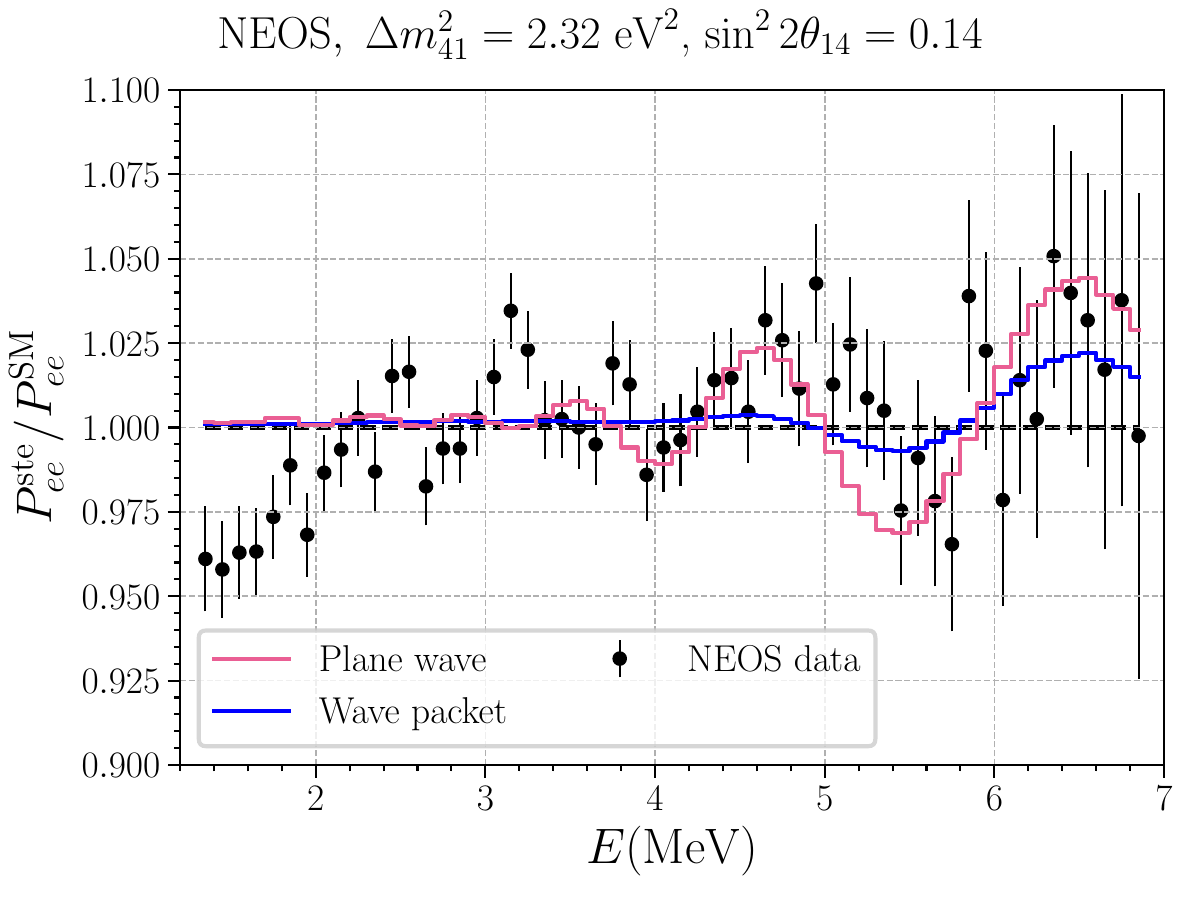}
    \caption{\textbf{\textit{Example of the effect in NEOS.}} A figure of the decoherence effect, for $\sigma_x = 2.1\times 10^{-4}\si\nm$, with the reactor antineutrino anomaly best-fit parameters~\cite{NEOS:2016wee}: $\Delta m^2_{41} = \SI{2.32}\eV^2$ and $\sin^22\theta_{14} = 0.14$.
    The $y$ axis represents the ratio between the 3+1 and the 3 expected events for the Daya Bay-NEOS joint analysis.}
    \label{fig:NEOSData}
\end{figure}

Our NEOS experiment analysis is based on the procedure in Ref.~\cite{Dentler:2017tkw,NEOS:2016wee} and using a $\chi^2$ function as its test statistic ($\mathcal{TS}^{\rm NEOS}(\theta_{14},\Delta m^2_{41},\vec \alpha)$).
As in Ref.~\cite{Dentler:2017tkw,NEOS:2016wee}, we have used the electron-antineutrino spectrum measured in the Daya Bay experiment~\cite{DayaBay:2018heb} as the source flux.
However, in our analysis, we also perform a combined fit of NEOS and Daya Bay, using the Huber-Mueller flux for both experiments modified by common nuisance parameters that accommodate for flux uncertainties.
\Cref{fig:NEOSData} shows the ratio between the expected events from a 3+1 model and from a 3 model at the NEOS baseline in this joint fit.
Here, the decoherence effect of the wave packet formalism is clearly manifest.

The PROSPECT data has also been analyzed following~\cite{Andriamirado2021}, where the detector is divided into different subsegments with different baselines, and using a $\chi^2$ function as its test statistic ($\mathcal{TS}^{\rm PROSPECT}(\theta_{14},\Delta m^2_{41})$) with a covariance provided by the experiment.
Since our PROSPECT analysis uses ratios, it is independent of the reactor flux model.
Finally, the combined test statistic used in the joint fit of Daya Bay, NEOS, and the PROSPECT is
\begin{align}
   \mathcal{TS}^{\rm Joint}(\theta_{14},\Delta m^2_{41}) =&\   \mathcal{TS}^{\rm PROSPECT}(\theta_{14},\Delta m^2_{41}) +\nonumber\\
   &\min_{\vec \alpha} \left[
   \mathcal{TS}^{\rm NEOS}(\theta_{14},\Delta m^2_{41},\vec \alpha) +\right.\nonumber\\
   &\left.\mathcal{TS}^{\rm DayaBay}(\theta_{14},\Delta m^2_{41},\vec \alpha)\right],
   \label{eq:test-statistic}
\end{align}obtained by adding the individual test statistics and minimizing over the correlated nuisance parameters.

\begin{figure}
    \centering
    \includegraphics[width = \linewidth]{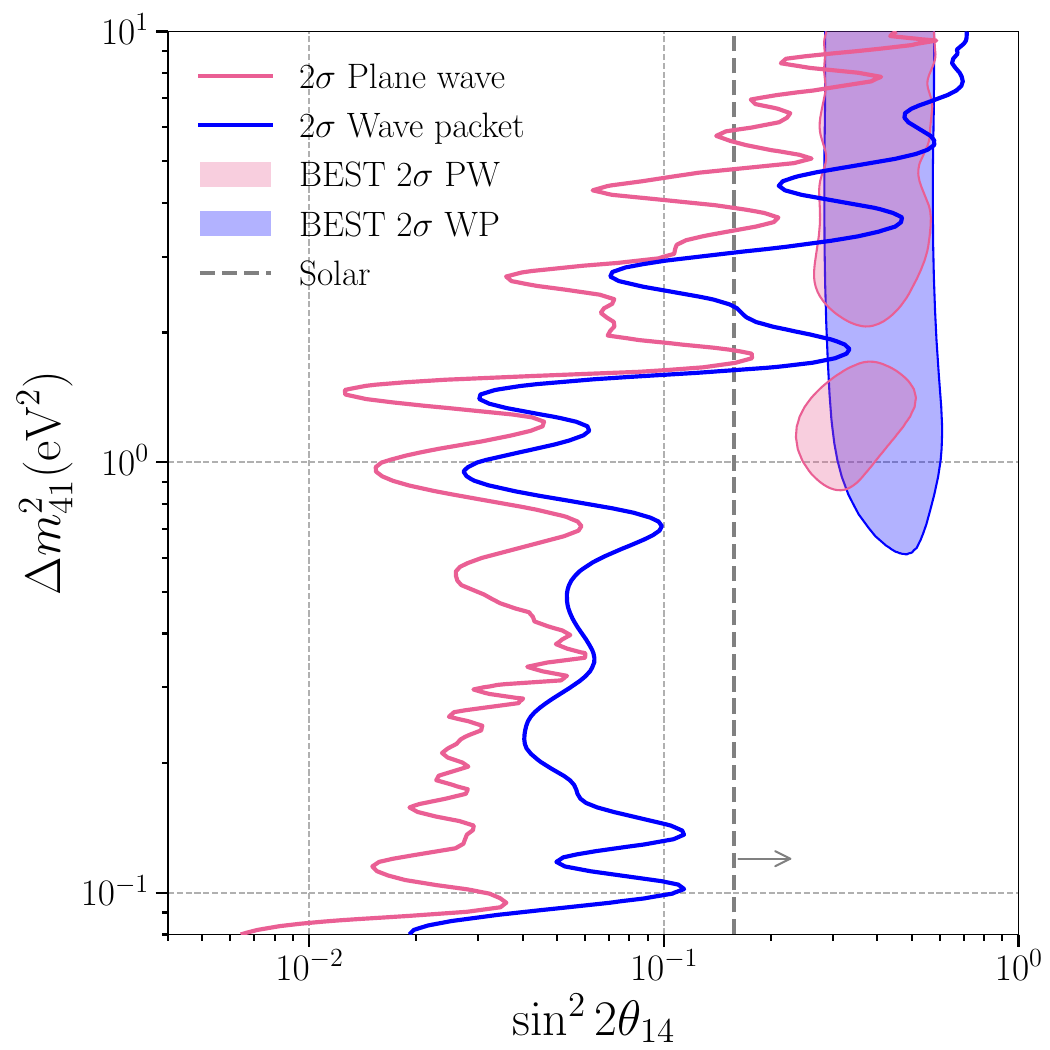}
    \caption{\textbf{\textit{Effect of finite wave-packet size on all the electron-neutrino disappearance experiments together.}} The solid pink and solid blue contours bound the exclusion region at two sigma for the plane wave approximation and wave packet formalism, respectively.
    The preferred region at two sigma for the BEST experiment is shaded for the plane wave approximation (pink) and the wave packet formalism (blue).
    All contours are obtained using $\sigma_x = 2.1\times 10^{-4}${\rm nm}.
    Notably, the region close to the global best fit point, $\Delta m^2_{41} \sim \SI{2}\eV$, is now allowed as well as a larger fraction of large mass-squared difference solutions.
    A gray dashed line marks the 2 sigma bounds from solar neutrino experiments~\cite{Berryman:2021yan,Goldhagen:2021kxe}.}
    \label{fig:allfit}
\end{figure}

We assume that the test statistic satisfies Wilk's theorem and draw the two-sigma exclusion contours in~\cref{fig:allfit}, which represent the main result of this paper.
Here, the solid pink line shows the exclusion regions at two sigma for the plane wave approximation, while the solid blue line is analogous with $\sigma_x = 2.1\times 10^{-4}\si\nm$ in the wave packet formalism. 
Finally, the BEST experiment has been fit with a two-point $\chi^2$ function, using the mean absorption rates for the inner and outer targets of the detector.
These rates can then be predicted as a function of the oscillation probability of the model.
The positive hint regions at two sigma by BEST are shown in~\cref{fig:allfit} as filled regions.
Again, pink is used for the plane wave approximation and blue for the wave packet formalism result. 
A couple of effects of decoherence can be noticed.
First, the suppression of oscillations connects the two separated regions around $\Delta m^2_{41} = \SI{2}\eV^2$, making both results compatible for values of $\Delta m^2_{41}$ that were excluded before.
Notice that in the lower $\Delta m^2_{41}$ part, the suppression of the event rate comes from a slow oscillation, and large values of $\sin^22\theta_{14}$ are needed to compensate for the decoherence effect.
Second, in the large $\Delta m^2_{41}$ region, the suppression of the event rate comes from the fast oscillations and therefore cannot be distinguished from a full decoherence effect. 

In this work, we have addressed only the tension between reactor and gallium experiments.
Aside from having a common energy range, this tension is specially interesting because both neutrino sources come from nuclear decay within a controlled environment, and systematic uncertainties are under control.
However, this is not the strongest tension in this energy range.
As can be seen in~\cref{fig:allfit}, recent solar analysis excludes a wide range of the allowed parameter space by reactor experiments, and are in tension with gallium experiments around three sigma.
We do not expect a finite wave packet size to affect this tension, and thus it is not addressed in this work.

\section{Conclusions}
In this paper, we studied the impact of the oscillation damping phenomena within the wave packet formalism in low-energy searches of sterile neutrinos.
Estimations of the wave packet sizes are currently larger than the experimental lower bound; however, these estimations are not without caveats.
We found that within the bounds for the wave packet sizes the effects are important in both the exclusion regions from nuclear reactors and the anomalous observations from radioactive sources measurements.
When setting the wave packet size at the current constraints, we find that the null observations using event ratios and the anomalous observations by BEST can be resolved.
The results become compatible not only at large values of $\Delta m^2_{41}$ but also at the region around $\Delta m^2_{41} = \SI{2}\eV^2$.
The work performed in this paper does not necessarily include additional new physics beyond a light sterile neutrino; instead, it highlights the importance of validating the plane wave approximation.

Our paper implies that further experimental work ought to be performed to understand decoherence effects in neutrino production and detection, and to constrain the size of the wave packet, since its impact is significant in sterile neutrino oscillations.
Additionally, we encourage calculations of the neutrino wave packet in the spirit of~\cite{Jones:2014sfa} for the relevant experimental contexts.
Our work additionally motivates the importance of understanding the reactor neutrino flux and the use of radioactive sources, whose fluxes are better predicted.
This is, there could be a scenario where the ratio experiments see null results in the presence of a sterile neutrino due to the effect mentioned, and the sterile neutrino could only be observed by comparing results to absolute flux predictions. Finally, our code is available \href{https://github.com/Harvard-Neutrino/DayaBaySterileDecoherence}{at this URL~\faGithub}. The experiments here analysed can be easily reproduced using arbitrary probabilities.

\section*{Acknowledgements}

We acknowledge Janet Conrad, Maria Concepcion Gonzalez-Garcia, and John Hardin.
We additionally thank Joachim Kopp and Alexei Smirnov for insightful discussions on wave packet size estimations and their role in neutrino decoherence. 
CAA is supported by the Faculty of Arts and Sciences of Harvard University, and the Alfred P. Sloan Foundation. 
TB and JS acknowledge financial support from the European ITN project H2020-MSCAITN-2019/860881-HIDDeN, the Spanish grants PID2019-108122GBC32, PID2019-105614GB-C21, and from the State Agency for Research of the Spanish Ministry of Science and Innovation through the ``Unit of Excellence María de Maeztu 2020-2023'' award to the Institute of Cosmos Sciences (CEX2019-000918-M).

\bibliography{wpdaya}

\begin{thebibliography}{76}%
\makeatletter
\providecommand \@ifxundefined [1]{%
 \@ifx{#1\undefined}
}%
\providecommand \@ifnum [1]{%
 \ifnum #1\expandafter \@firstoftwo
 \else \expandafter \@secondoftwo
 \fi
}%
\providecommand \@ifx [1]{%
 \ifx #1\expandafter \@firstoftwo
 \else \expandafter \@secondoftwo
 \fi
}%
\providecommand \natexlab [1]{#1}%
\providecommand \enquote  [1]{``#1''}%
\providecommand \bibnamefont  [1]{#1}%
\providecommand \bibfnamefont [1]{#1}%
\providecommand \citenamefont [1]{#1}%
\providecommand \href@noop [0]{\@secondoftwo}%
\providecommand \href [0]{\begingroup \@sanitize@url \@href}%
\providecommand \@href[1]{\@@startlink{#1}\@@href}%
\providecommand \@@href[1]{\endgroup#1\@@endlink}%
\providecommand \@sanitize@url [0]{\catcode `\\12\catcode `\$12\catcode
  `\&12\catcode `\#12\catcode `\^12\catcode `\_12\catcode `\%12\relax}%
\providecommand \@@startlink[1]{}%
\providecommand \@@endlink[0]{}%
\providecommand \url  [0]{\begingroup\@sanitize@url \@url }%
\providecommand \@url [1]{\endgroup\@href {#1}{\urlprefix }}%
\providecommand \urlprefix  [0]{URL }%
\providecommand \Eprint [0]{\href }%
\providecommand \doibase [0]{http://dx.doi.org/}%
\providecommand \selectlanguage [0]{\@gobble}%
\providecommand \bibinfo  [0]{\@secondoftwo}%
\providecommand \bibfield  [0]{\@secondoftwo}%
\providecommand \translation [1]{[#1]}%
\providecommand \BibitemOpen [0]{}%
\providecommand \bibitemStop [0]{}%
\providecommand \bibitemNoStop [0]{.\EOS\space}%
\providecommand \EOS [0]{\spacefactor3000\relax}%
\providecommand \BibitemShut  [1]{\csname bibitem#1\endcsname}%
\let\auto@bib@innerbib\@empty
\bibitem [{\citenamefont {Athanassopoulos}\ \emph {et~al.}(1996)\citenamefont
  {Athanassopoulos} \emph {et~al.}}]{LSND:1996ubh}%
  \BibitemOpen
  \bibfield  {author} {\bibinfo {author} {\bibfnamefont {C.}~\bibnamefont
  {Athanassopoulos}} \emph {et~al.} (\bibinfo {collaboration} {LSND}),\
  }\bibfield  {title} {\enquote {\bibinfo {title} {{Evidence for
  anti-muon-neutrino ---\ensuremath{>} anti-electron-neutrino oscillations from
  the LSND experiment at LAMPF}},}\ }\href {\doibase
  10.1103/PhysRevLett.77.3082} {\bibfield  {journal} {\bibinfo  {journal}
  {Phys. Rev. Lett.}\ }\textbf {\bibinfo {volume} {77}},\ \bibinfo {pages}
  {3082--3085} (\bibinfo {year} {1996})},\ \Eprint
  {http://arxiv.org/abs/nucl-ex/9605003} {arXiv:nucl-ex/9605003} \BibitemShut
  {NoStop}%
\bibitem [{\citenamefont {Aguilar-Arevalo}\ \emph {et~al.}(2001)\citenamefont
  {Aguilar-Arevalo} \emph {et~al.}}]{LSND:2001aii}%
  \BibitemOpen
  \bibfield  {author} {\bibinfo {author} {\bibfnamefont {A.}~\bibnamefont
  {Aguilar-Arevalo}} \emph {et~al.} (\bibinfo {collaboration} {LSND}),\
  }\bibfield  {title} {\enquote {\bibinfo {title} {{Evidence for neutrino
  oscillations from the observation of $\bar{\nu}_e$ appearance in a
  $\bar{\nu}_\mu$ beam}},}\ }\href {\doibase 10.1103/PhysRevD.64.112007}
  {\bibfield  {journal} {\bibinfo  {journal} {Phys. Rev. D}\ }\textbf {\bibinfo
  {volume} {64}},\ \bibinfo {pages} {112007} (\bibinfo {year} {2001})},\
  \Eprint {http://arxiv.org/abs/hep-ex/0104049} {arXiv:hep-ex/0104049}
  \BibitemShut {NoStop}%
\bibitem [{\citenamefont {Bahcall}\ \emph {et~al.}(1995)\citenamefont
  {Bahcall}, \citenamefont {Krastev},\ and\ \citenamefont
  {Lisi}}]{Bahcall:1994bq}%
  \BibitemOpen
  \bibfield  {author} {\bibinfo {author} {\bibfnamefont {John~N.}\ \bibnamefont
  {Bahcall}}, \bibinfo {author} {\bibfnamefont {P.~I.}\ \bibnamefont
  {Krastev}}, \ and\ \bibinfo {author} {\bibfnamefont {E.}~\bibnamefont
  {Lisi}},\ }\bibfield  {title} {\enquote {\bibinfo {title} {{Limits on
  electron-neutrino oscillations from the GALLEX Cr-51 source experiment}},}\
  }\href {\doibase 10.1016/0370-2693(95)00111-W} {\bibfield  {journal}
  {\bibinfo  {journal} {Phys. Lett. B}\ }\textbf {\bibinfo {volume} {348}},\
  \bibinfo {pages} {121--123} (\bibinfo {year} {1995})},\ \Eprint
  {http://arxiv.org/abs/hep-ph/9411414} {arXiv:hep-ph/9411414} \BibitemShut
  {NoStop}%
\bibitem [{\citenamefont {Bahcall}(1997)}]{Bahcall1997}%
  \BibitemOpen
  \bibfield  {author} {\bibinfo {author} {\bibfnamefont {John~N.}\ \bibnamefont
  {Bahcall}},\ }\bibfield  {title} {\enquote {\bibinfo {title} {Gallium solar
  neutrino experiments: Absorption cross sections, neutrino spectra, and
  predicted event rates},}\ }\href {\doibase 10.1103/PhysRevC.56.3391}
  {\bibfield  {journal} {\bibinfo  {journal} {Phys. Rev. C}\ }\textbf {\bibinfo
  {volume} {56}},\ \bibinfo {pages} {3391--3409} (\bibinfo {year}
  {1997})}\BibitemShut {NoStop}%
\bibitem [{\citenamefont {Abdurashitov}\ \emph {et~al.}(1999)\citenamefont
  {Abdurashitov} \emph {et~al.}}]{SAGE:1998fvr}%
  \BibitemOpen
  \bibfield  {author} {\bibinfo {author} {\bibfnamefont {J.~N.}\ \bibnamefont
  {Abdurashitov}} \emph {et~al.} (\bibinfo {collaboration} {SAGE}),\ }\bibfield
   {title} {\enquote {\bibinfo {title} {{Measurement of the response of the
  Russian-American gallium experiment to neutrinos from a Cr-51 source}},}\
  }\href {\doibase 10.1103/PhysRevC.59.2246} {\bibfield  {journal} {\bibinfo
  {journal} {Phys. Rev. C}\ }\textbf {\bibinfo {volume} {59}},\ \bibinfo
  {pages} {2246--2263} (\bibinfo {year} {1999})},\ \Eprint
  {http://arxiv.org/abs/hep-ph/9803418} {arXiv:hep-ph/9803418} \BibitemShut
  {NoStop}%
\bibitem [{\citenamefont {Barinov}\ \emph {et~al.}(2021)\citenamefont {Barinov}
  \emph {et~al.}}]{Barinov:2021asz}%
  \BibitemOpen
  \bibfield  {author} {\bibinfo {author} {\bibfnamefont {V.~V.}\ \bibnamefont
  {Barinov}} \emph {et~al.},\ }\bibfield  {title} {\enquote {\bibinfo {title}
  {{Results from the Baksan Experiment on Sterile Transitions (BEST)}},}\
  }\href@noop {} {\  (\bibinfo {year} {2021})},\ \Eprint
  {http://arxiv.org/abs/2109.11482} {arXiv:2109.11482 [nucl-ex]} \BibitemShut
  {NoStop}%
\bibitem [{\citenamefont {Giunti}\ and\ \citenamefont
  {Laveder}(2011)}]{Giunti:2010zu}%
  \BibitemOpen
  \bibfield  {author} {\bibinfo {author} {\bibfnamefont {Carlo}\ \bibnamefont
  {Giunti}}\ and\ \bibinfo {author} {\bibfnamefont {Marco}\ \bibnamefont
  {Laveder}},\ }\bibfield  {title} {\enquote {\bibinfo {title} {{Statistical
  Significance of the Gallium Anomaly}},}\ }\href {\doibase
  10.1103/PhysRevC.83.065504} {\bibfield  {journal} {\bibinfo  {journal} {Phys.
  Rev. C}\ }\textbf {\bibinfo {volume} {83}},\ \bibinfo {pages} {065504}
  (\bibinfo {year} {2011})},\ \Eprint {http://arxiv.org/abs/1006.3244}
  {arXiv:1006.3244 [hep-ph]} \BibitemShut {NoStop}%
\bibitem [{\citenamefont {Abdurashitov}\ \emph {et~al.}(2006)\citenamefont
  {Abdurashitov} \emph {et~al.}}]{Abdurashitov:2005tb}%
  \BibitemOpen
  \bibfield  {author} {\bibinfo {author} {\bibfnamefont {J.~N.}\ \bibnamefont
  {Abdurashitov}} \emph {et~al.},\ }\bibfield  {title} {\enquote {\bibinfo
  {title} {{Measurement of the response of a Ga solar neutrino experiment to
  neutrinos from an Ar-37 source}},}\ }\href {\doibase
  10.1103/PhysRevC.73.045805} {\bibfield  {journal} {\bibinfo  {journal} {Phys.
  Rev. C}\ }\textbf {\bibinfo {volume} {73}},\ \bibinfo {pages} {045805}
  (\bibinfo {year} {2006})},\ \Eprint {http://arxiv.org/abs/nucl-ex/0512041}
  {arXiv:nucl-ex/0512041} \BibitemShut {NoStop}%
\bibitem [{\citenamefont {Mention}\ \emph {et~al.}(2011)\citenamefont
  {Mention}, \citenamefont {Fechner}, \citenamefont {Lasserre}, \citenamefont
  {Mueller}, \citenamefont {Lhuillier}, \citenamefont {Cribier},\ and\
  \citenamefont {Letourneau}}]{Mention:2011rk}%
  \BibitemOpen
  \bibfield  {author} {\bibinfo {author} {\bibfnamefont {G.}~\bibnamefont
  {Mention}}, \bibinfo {author} {\bibfnamefont {M.}~\bibnamefont {Fechner}},
  \bibinfo {author} {\bibfnamefont {Th.}\ \bibnamefont {Lasserre}}, \bibinfo
  {author} {\bibfnamefont {Th.~A.}\ \bibnamefont {Mueller}}, \bibinfo {author}
  {\bibfnamefont {D.}~\bibnamefont {Lhuillier}}, \bibinfo {author}
  {\bibfnamefont {M.}~\bibnamefont {Cribier}}, \ and\ \bibinfo {author}
  {\bibfnamefont {A.}~\bibnamefont {Letourneau}},\ }\bibfield  {title}
  {\enquote {\bibinfo {title} {{The Reactor Antineutrino Anomaly}},}\ }\href
  {\doibase 10.1103/PhysRevD.83.073006} {\bibfield  {journal} {\bibinfo
  {journal} {Phys. Rev. D}\ }\textbf {\bibinfo {volume} {83}},\ \bibinfo
  {pages} {073006} (\bibinfo {year} {2011})},\ \Eprint
  {http://arxiv.org/abs/1101.2755} {arXiv:1101.2755 [hep-ex]} \BibitemShut
  {NoStop}%
\bibitem [{\citenamefont {An}\ \emph {et~al.}(2016)\citenamefont {An} \emph
  {et~al.}}]{DayaBay:2016qvc}%
  \BibitemOpen
  \bibfield  {author} {\bibinfo {author} {\bibfnamefont {Feng~Peng}\
  \bibnamefont {An}} \emph {et~al.} (\bibinfo {collaboration} {Daya Bay}),\
  }\bibfield  {title} {\enquote {\bibinfo {title} {{Improved Search for a Light
  Sterile Neutrino with the Full Configuration of the Daya Bay Experiment}},}\
  }\href {\doibase 10.1103/PhysRevLett.117.151802} {\bibfield  {journal}
  {\bibinfo  {journal} {Phys. Rev. Lett.}\ }\textbf {\bibinfo {volume} {117}},\
  \bibinfo {pages} {151802} (\bibinfo {year} {2016})},\ \Eprint
  {http://arxiv.org/abs/1607.01174} {arXiv:1607.01174 [hep-ex]} \BibitemShut
  {NoStop}%
\bibitem [{\citenamefont {An}\ \emph {et~al.}(2017{\natexlab{a}})\citenamefont
  {An} \emph {et~al.}}]{DayaBay:2016ggj}%
  \BibitemOpen
  \bibfield  {author} {\bibinfo {author} {\bibfnamefont {Feng~Peng}\
  \bibnamefont {An}} \emph {et~al.} (\bibinfo {collaboration} {Daya Bay}),\
  }\bibfield  {title} {\enquote {\bibinfo {title} {{Measurement of electron
  antineutrino oscillation based on 1230 days of operation of the Daya Bay
  experiment}},}\ }\href {\doibase 10.1103/PhysRevD.95.072006} {\bibfield
  {journal} {\bibinfo  {journal} {Phys. Rev. D}\ }\textbf {\bibinfo {volume}
  {95}},\ \bibinfo {pages} {072006} (\bibinfo {year} {2017}{\natexlab{a}})},\
  \Eprint {http://arxiv.org/abs/1610.04802} {arXiv:1610.04802 [hep-ex]}
  \BibitemShut {NoStop}%
\bibitem [{\citenamefont {Ko}\ \emph {et~al.}(2017)\citenamefont {Ko} \emph
  {et~al.}}]{NEOS:2016wee}%
  \BibitemOpen
  \bibfield  {author} {\bibinfo {author} {\bibfnamefont {Y.~J.}\ \bibnamefont
  {Ko}} \emph {et~al.} (\bibinfo {collaboration} {NEOS}),\ }\bibfield  {title}
  {\enquote {\bibinfo {title} {{Sterile Neutrino Search at the NEOS
  Experiment}},}\ }\href {\doibase 10.1103/PhysRevLett.118.121802} {\bibfield
  {journal} {\bibinfo  {journal} {Phys. Rev. Lett.}\ }\textbf {\bibinfo
  {volume} {118}},\ \bibinfo {pages} {121802} (\bibinfo {year} {2017})},\
  \Eprint {http://arxiv.org/abs/1610.05134} {arXiv:1610.05134 [hep-ex]}
  \BibitemShut {NoStop}%
\bibitem [{\citenamefont {Almazan}\ \emph {et~al.}(2018)\citenamefont {Almazan}
  \emph {et~al.}}]{Stereo66}%
  \BibitemOpen
  \bibfield  {author} {\bibinfo {author} {\bibfnamefont {H.}~\bibnamefont
  {Almazan}} \emph {et~al.} (\bibinfo {collaboration} {STEREO}),\ }\bibfield
  {title} {\enquote {\bibinfo {title} {{``Sterile Neutrino Constraints from the
  STEREO Experiment with 66 Days of Reactor-On Data''}},}\ }\href {\doibase
  10.1103/PhysRevLett.121.161801} {\bibfield  {journal} {\bibinfo  {journal}
  {Phys. Rev. Lett.}\ }\textbf {\bibinfo {volume} {121}},\ \bibinfo {pages}
  {161801} (\bibinfo {year} {2018})},\ \Eprint
  {http://arxiv.org/abs/1806.02096} {arXiv:1806.02096 [hep-ex]} \BibitemShut
  {NoStop}%
\bibitem [{\citenamefont {Manzanillas}(2018)}]{solid}%
  \BibitemOpen
  \bibfield  {author} {\bibinfo {author} {\bibfnamefont {Luis}\ \bibnamefont
  {Manzanillas}},\ }\bibfield  {title} {\enquote {\bibinfo {title}
  {{Performance of the SoLid Reactor Neutrino Detector}},}\ }in\ \href@noop {}
  {\emph {\bibinfo {booktitle} {{39th International Conference on High Energy
  Physics (ICHEP 2018) Seoul, Gangnam-Gu, Korea, Republic of, July 4-11,
  2018}}}}\ (\bibinfo {year} {2018})\ \Eprint {http://arxiv.org/abs/1811.05694}
  {arXiv:1811.05694 [physics.ins-det]} \BibitemShut {NoStop}%
\bibitem [{\citenamefont {Serebrov}\ \emph {et~al.}(2021)\citenamefont
  {Serebrov}, \citenamefont {Samoilov}, \citenamefont {Ivochkin}, \citenamefont
  {Fomin}, \citenamefont {Zinoviev}, \citenamefont {Neustroev}, \citenamefont
  {Golovtsov}, \citenamefont {Volkov}, \citenamefont {Chernyj}, \citenamefont
  {Zherebtsov},\ and\ \citenamefont {et~al.}}]{neutrino42021}%
  \BibitemOpen
  \bibfield  {author} {\bibinfo {author} {\bibfnamefont {A.~P.}\ \bibnamefont
  {Serebrov}}, \bibinfo {author} {\bibfnamefont {R.~M.}\ \bibnamefont
  {Samoilov}}, \bibinfo {author} {\bibfnamefont {V.~G.}\ \bibnamefont
  {Ivochkin}}, \bibinfo {author} {\bibfnamefont {A.~K.}\ \bibnamefont {Fomin}},
  \bibinfo {author} {\bibfnamefont {V.~G.}\ \bibnamefont {Zinoviev}}, \bibinfo
  {author} {\bibfnamefont {P.~V.}\ \bibnamefont {Neustroev}}, \bibinfo {author}
  {\bibfnamefont {V.~L.}\ \bibnamefont {Golovtsov}}, \bibinfo {author}
  {\bibfnamefont {S.~S.}\ \bibnamefont {Volkov}}, \bibinfo {author}
  {\bibfnamefont {A.~V.}\ \bibnamefont {Chernyj}}, \bibinfo {author}
  {\bibfnamefont {O.~M.}\ \bibnamefont {Zherebtsov}}, \ and\ \bibinfo {author}
  {\bibnamefont {et~al.}},\ }\bibfield  {title} {\enquote {\bibinfo {title}
  {Search for sterile neutrinos with the neutrino-4 experiment and measurement
  results},}\ }\href {\doibase 10.1103/physrevd.104.032003} {\bibfield
  {journal} {\bibinfo  {journal} {Physical Review D}\ }\textbf {\bibinfo
  {volume} {104}} (\bibinfo {year} {2021}),\
  10.1103/physrevd.104.032003}\BibitemShut {NoStop}%
\bibitem [{\citenamefont {Choi}\ \emph {et~al.}(2020)\citenamefont {Choi} \emph
  {et~al.}}]{PhysRevLett.125.191801}%
  \BibitemOpen
  \bibfield  {author} {\bibinfo {author} {\bibfnamefont {J.~H.}\ \bibnamefont
  {Choi}} \emph {et~al.} (\bibinfo {collaboration} {RENO Collaboration}),\
  }\bibfield  {title} {\enquote {\bibinfo {title} {Search for sub-ev sterile
  neutrinos at reno},}\ }\href {\doibase 10.1103/PhysRevLett.125.191801}
  {\bibfield  {journal} {\bibinfo  {journal} {Phys. Rev. Lett.}\ }\textbf
  {\bibinfo {volume} {125}},\ \bibinfo {pages} {191801} (\bibinfo {year}
  {2020})}\BibitemShut {NoStop}%
\bibitem [{\citenamefont {Andriamirado}\ \emph {et~al.}(2021)\citenamefont
  {Andriamirado} \emph {et~al.}}]{Andriamirado2021}%
  \BibitemOpen
  \bibfield  {author} {\bibinfo {author} {\bibfnamefont {M.}~\bibnamefont
  {Andriamirado}} \emph {et~al.},\ }\bibfield  {title} {\enquote {\bibinfo
  {title} {Improved short-baseline neutrino oscillation search and energy
  spectrum measurement with the {PROSPECT} experiment at {HFIR}},}\ }\href
  {\doibase 10.1103/physrevd.103.032001} {\bibfield  {journal} {\bibinfo
  {journal} {Phys. Rev. D 103, 032001 (2021)}\ }\textbf {\bibinfo {volume}
  {103}},\ \bibinfo {pages} {032001} (\bibinfo {year} {2021})},\ \Eprint
  {http://arxiv.org/abs/https://arxiv.org/abs/2006.11210}
  {arXiv:https://arxiv.org/abs/2006.11210 [hep-ex]} \BibitemShut {NoStop}%
\bibitem [{\citenamefont {Barinov}\ and\ \citenamefont
  {Gorbunov}(2021)}]{Barinov:2021mjj}%
  \BibitemOpen
  \bibfield  {author} {\bibinfo {author} {\bibfnamefont {Vladislav}\
  \bibnamefont {Barinov}}\ and\ \bibinfo {author} {\bibfnamefont {Dmitry}\
  \bibnamefont {Gorbunov}},\ }\bibfield  {title} {\enquote {\bibinfo {title}
  {{BEST Impact on Sterile Neutrino Hypothesis}},}\ }\href@noop {} {\
  (\bibinfo {year} {2021})},\ \Eprint {http://arxiv.org/abs/2109.14654}
  {arXiv:2109.14654 [hep-ph]} \BibitemShut {NoStop}%
\bibitem [{\citenamefont {Alekseev}\ \emph {et~al.}(2018)\citenamefont
  {Alekseev} \emph {et~al.}}]{DANSS:2018fnn}%
  \BibitemOpen
  \bibfield  {author} {\bibinfo {author} {\bibfnamefont {I}~\bibnamefont
  {Alekseev}} \emph {et~al.} (\bibinfo {collaboration} {DANSS}),\ }\bibfield
  {title} {\enquote {\bibinfo {title} {{Search for sterile neutrinos at the
  DANSS experiment}},}\ }\href {\doibase 10.1016/j.physletb.2018.10.038}
  {\bibfield  {journal} {\bibinfo  {journal} {Phys. Lett. B}\ }\textbf
  {\bibinfo {volume} {787}},\ \bibinfo {pages} {56--63} (\bibinfo {year}
  {2018})},\ \Eprint {http://arxiv.org/abs/1804.04046} {arXiv:1804.04046
  [hep-ex]} \BibitemShut {NoStop}%
\bibitem [{\citenamefont {Danilov}\ and\ \citenamefont
  {Skrobova}(2021)}]{Danilov:2021oop}%
  \BibitemOpen
  \bibfield  {author} {\bibinfo {author} {\bibfnamefont {Mikhail}\ \bibnamefont
  {Danilov}}\ and\ \bibinfo {author} {\bibfnamefont {Nataliya}\ \bibnamefont
  {Skrobova}} (\bibinfo {collaboration} {DANSS}),\ }\bibfield  {title}
  {\enquote {\bibinfo {title} {{New results from the DANSS experiment}},}\ }in\
  \href@noop {} {\emph {\bibinfo {booktitle} {{European Physical Society
  Conference on High Energy Physics 2021}}}}\ (\bibinfo {year} {2021})\ \Eprint
  {http://arxiv.org/abs/2112.13413} {arXiv:2112.13413 [hep-ex]} \BibitemShut
  {NoStop}%
\bibitem [{\citenamefont {Berryman}\ \emph {et~al.}(2022)\citenamefont
  {Berryman}, \citenamefont {Coloma}, \citenamefont {Huber}, \citenamefont
  {Schwetz},\ and\ \citenamefont {Zhou}}]{Berryman:2021yan}%
  \BibitemOpen
  \bibfield  {author} {\bibinfo {author} {\bibfnamefont {Jeffrey~M.}\
  \bibnamefont {Berryman}}, \bibinfo {author} {\bibfnamefont {Pilar}\
  \bibnamefont {Coloma}}, \bibinfo {author} {\bibfnamefont {Patrick}\
  \bibnamefont {Huber}}, \bibinfo {author} {\bibfnamefont {Thomas}\
  \bibnamefont {Schwetz}}, \ and\ \bibinfo {author} {\bibfnamefont {Albert}\
  \bibnamefont {Zhou}},\ }\bibfield  {title} {\enquote {\bibinfo {title}
  {{Statistical significance of the sterile-neutrino hypothesis in the context
  of reactor and gallium data}},}\ }\href {\doibase 10.1007/JHEP02(2022)055}
  {\bibfield  {journal} {\bibinfo  {journal} {JHEP}\ }\textbf {\bibinfo
  {volume} {02}},\ \bibinfo {pages} {055} (\bibinfo {year} {2022})},\ \Eprint
  {http://arxiv.org/abs/2111.12530} {arXiv:2111.12530 [hep-ph]} \BibitemShut
  {NoStop}%
\bibitem [{\citenamefont {Goldhagen}\ \emph {et~al.}(2022)\citenamefont
  {Goldhagen}, \citenamefont {Maltoni}, \citenamefont {Reichard},\ and\
  \citenamefont {Schwetz}}]{Goldhagen:2021kxe}%
  \BibitemOpen
  \bibfield  {author} {\bibinfo {author} {\bibfnamefont {Kim}\ \bibnamefont
  {Goldhagen}}, \bibinfo {author} {\bibfnamefont {Michele}\ \bibnamefont
  {Maltoni}}, \bibinfo {author} {\bibfnamefont {Shayne~E.}\ \bibnamefont
  {Reichard}}, \ and\ \bibinfo {author} {\bibfnamefont {Thomas}\ \bibnamefont
  {Schwetz}},\ }\bibfield  {title} {\enquote {\bibinfo {title} {{Testing
  sterile neutrino mixing with present and future solar neutrino data}},}\
  }\href {\doibase 10.1140/epjc/s10052-022-10052-2} {\bibfield  {journal}
  {\bibinfo  {journal} {Eur. Phys. J. C}\ }\textbf {\bibinfo {volume} {82}},\
  \bibinfo {pages} {116} (\bibinfo {year} {2022})},\ \Eprint
  {http://arxiv.org/abs/2109.14898} {arXiv:2109.14898 [hep-ph]} \BibitemShut
  {NoStop}%
\bibitem [{\citenamefont {Aguilar-Arevalo}\ \emph {et~al.}(2013)\citenamefont
  {Aguilar-Arevalo} \emph {et~al.}}]{MiniBooNE:2013uba}%
  \BibitemOpen
  \bibfield  {author} {\bibinfo {author} {\bibfnamefont {A.~A.}\ \bibnamefont
  {Aguilar-Arevalo}} \emph {et~al.} (\bibinfo {collaboration} {MiniBooNE}),\
  }\bibfield  {title} {\enquote {\bibinfo {title} {{Improved Search for $\bar
  \nu_\mu \rightarrow \bar \nu_e$ Oscillations in the MiniBooNE Experiment}},}\
  }\href {\doibase 10.1103/PhysRevLett.110.161801} {\bibfield  {journal}
  {\bibinfo  {journal} {Phys. Rev. Lett.}\ }\textbf {\bibinfo {volume} {110}},\
  \bibinfo {pages} {161801} (\bibinfo {year} {2013})},\ \Eprint
  {http://arxiv.org/abs/1303.2588} {arXiv:1303.2588 [hep-ex]} \BibitemShut
  {NoStop}%
\bibitem [{\citenamefont {Aguilar-Arevalo}\ \emph {et~al.}(2018)\citenamefont
  {Aguilar-Arevalo} \emph {et~al.}}]{MiniBooNE:2018esg}%
  \BibitemOpen
  \bibfield  {author} {\bibinfo {author} {\bibfnamefont {A.~A.}\ \bibnamefont
  {Aguilar-Arevalo}} \emph {et~al.} (\bibinfo {collaboration} {MiniBooNE}),\
  }\bibfield  {title} {\enquote {\bibinfo {title} {{Significant Excess of
  ElectronLike Events in the MiniBooNE Short-Baseline Neutrino Experiment}},}\
  }\href {\doibase 10.1103/PhysRevLett.121.221801} {\bibfield  {journal}
  {\bibinfo  {journal} {Phys. Rev. Lett.}\ }\textbf {\bibinfo {volume} {121}},\
  \bibinfo {pages} {221801} (\bibinfo {year} {2018})},\ \Eprint
  {http://arxiv.org/abs/1805.12028} {arXiv:1805.12028 [hep-ex]} \BibitemShut
  {NoStop}%
\bibitem [{\citenamefont {Abratenko}\ \emph
  {et~al.}(2021{\natexlab{a}})\citenamefont {Abratenko} \emph
  {et~al.}}]{MicroBooNE:2021jwr}%
  \BibitemOpen
  \bibfield  {author} {\bibinfo {author} {\bibfnamefont {P.}~\bibnamefont
  {Abratenko}} \emph {et~al.} (\bibinfo {collaboration} {MicroBooNE}),\
  }\bibfield  {title} {\enquote {\bibinfo {title} {{Search for an anomalous
  excess of charged-current quasi-elastic $\nu_e$ interactions with the
  MicroBooNE experiment using Deep-Learning-based reconstruction}},}\
  }\href@noop {} {\  (\bibinfo {year} {2021}{\natexlab{a}})},\ \Eprint
  {http://arxiv.org/abs/2110.14080} {arXiv:2110.14080 [hep-ex]} \BibitemShut
  {NoStop}%
\bibitem [{\citenamefont {Abratenko}\ \emph
  {et~al.}(2021{\natexlab{b}})\citenamefont {Abratenko} \emph
  {et~al.}}]{MicroBooNE:2021nxr}%
  \BibitemOpen
  \bibfield  {author} {\bibinfo {author} {\bibfnamefont {P.}~\bibnamefont
  {Abratenko}} \emph {et~al.} (\bibinfo {collaboration} {MicroBooNE}),\
  }\bibfield  {title} {\enquote {\bibinfo {title} {{Search for an anomalous
  excess of inclusive charged-current $\nu_e$ interactions in the MicroBooNE
  experiment using Wire-Cell reconstruction}},}\ }\href@noop {} {\  (\bibinfo
  {year} {2021}{\natexlab{b}})},\ \Eprint {http://arxiv.org/abs/2110.13978}
  {arXiv:2110.13978 [hep-ex]} \BibitemShut {NoStop}%
\bibitem [{\citenamefont {Abratenko}\ \emph
  {et~al.}(2021{\natexlab{c}})\citenamefont {Abratenko} \emph
  {et~al.}}]{MicroBooNE:2021rmx}%
  \BibitemOpen
  \bibfield  {author} {\bibinfo {author} {\bibfnamefont {P.}~\bibnamefont
  {Abratenko}} \emph {et~al.} (\bibinfo {collaboration} {MicroBooNE}),\
  }\bibfield  {title} {\enquote {\bibinfo {title} {{Search for an Excess of
  Electron Neutrino Interactions in MicroBooNE Using Multiple Final State
  Topologies}},}\ }\href@noop {} {\  (\bibinfo {year} {2021}{\natexlab{c}})},\
  \Eprint {http://arxiv.org/abs/2110.14054} {arXiv:2110.14054 [hep-ex]}
  \BibitemShut {NoStop}%
\bibitem [{\citenamefont {Abratenko}\ \emph
  {et~al.}(2021{\natexlab{d}})\citenamefont {Abratenko} \emph
  {et~al.}}]{MicroBooNE:2021sne}%
  \BibitemOpen
  \bibfield  {author} {\bibinfo {author} {\bibfnamefont {P.}~\bibnamefont
  {Abratenko}} \emph {et~al.} (\bibinfo {collaboration} {MicroBooNE}),\
  }\bibfield  {title} {\enquote {\bibinfo {title} {{Search for an anomalous
  excess of charged-current $\nu_e$ interactions without pions in the final
  state with the MicroBooNE experiment}},}\ }\href@noop {} {\  (\bibinfo {year}
  {2021}{\natexlab{d}})},\ \Eprint {http://arxiv.org/abs/2110.14065}
  {arXiv:2110.14065 [hep-ex]} \BibitemShut {NoStop}%
\bibitem [{\citenamefont {Denton}(2021)}]{Denton:2021czb}%
  \BibitemOpen
  \bibfield  {author} {\bibinfo {author} {\bibfnamefont {Peter~B.}\
  \bibnamefont {Denton}},\ }\bibfield  {title} {\enquote {\bibinfo {title}
  {{Sterile Neutrino Searches with MicroBooNE: Electron Neutrino
  Disappearance}},}\ }\href@noop {} {\  (\bibinfo {year} {2021})},\ \Eprint
  {http://arxiv.org/abs/2111.05793} {arXiv:2111.05793 [hep-ph]} \BibitemShut
  {NoStop}%
\bibitem [{\citenamefont {Arg\"uelles}\ \emph {et~al.}(2021)\citenamefont
  {Arg\"uelles}, \citenamefont {Esteban}, \citenamefont {Hostert},
  \citenamefont {Kelly}, \citenamefont {Kopp}, \citenamefont {Machado},
  \citenamefont {Martinez-Soler},\ and\ \citenamefont
  {Perez-Gonzalez}}]{Arguelles:2021meu}%
  \BibitemOpen
  \bibfield  {author} {\bibinfo {author} {\bibfnamefont {C.~A.}\ \bibnamefont
  {Arg\"uelles}}, \bibinfo {author} {\bibfnamefont {I.}~\bibnamefont
  {Esteban}}, \bibinfo {author} {\bibfnamefont {M.}~\bibnamefont {Hostert}},
  \bibinfo {author} {\bibfnamefont {K.~J.}\ \bibnamefont {Kelly}}, \bibinfo
  {author} {\bibfnamefont {J.}~\bibnamefont {Kopp}}, \bibinfo {author}
  {\bibfnamefont {P.~A.~N.}\ \bibnamefont {Machado}}, \bibinfo {author}
  {\bibfnamefont {I.}~\bibnamefont {Martinez-Soler}}, \ and\ \bibinfo {author}
  {\bibfnamefont {Y.~F.}\ \bibnamefont {Perez-Gonzalez}},\ }\bibfield  {title}
  {\enquote {\bibinfo {title} {{MicroBooNE and the $\nu_e$ Interpretation of
  the MiniBooNE Low-Energy Excess}},}\ }\href@noop {} {\  (\bibinfo {year}
  {2021})},\ \Eprint {http://arxiv.org/abs/2111.10359} {arXiv:2111.10359
  [hep-ph]} \BibitemShut {NoStop}%
\bibitem [{\citenamefont {Aguilar-Arevalo}\ \emph {et~al.}(2022)\citenamefont
  {Aguilar-Arevalo} \emph {et~al.}}]{MiniBooNE:2022emn}%
  \BibitemOpen
  \bibfield  {author} {\bibinfo {author} {\bibfnamefont {A.~A.}\ \bibnamefont
  {Aguilar-Arevalo}} \emph {et~al.} (\bibinfo {collaboration} {MiniBooNE}),\
  }\bibfield  {title} {\enquote {\bibinfo {title} {{MiniBooNE and MicroBooNE
  Joint Fit to a 3+1 Sterile Neutrino Scenario}},}\ }\href@noop {} {\
  (\bibinfo {year} {2022})},\ \Eprint {http://arxiv.org/abs/2201.01724}
  {arXiv:2201.01724 [hep-ex]} \BibitemShut {NoStop}%
\bibitem [{\citenamefont {Adamson}\ \emph {et~al.}(2019)\citenamefont {Adamson}
  \emph {et~al.}}]{MINOS:2017cae}%
  \BibitemOpen
  \bibfield  {author} {\bibinfo {author} {\bibfnamefont {P.}~\bibnamefont
  {Adamson}} \emph {et~al.} (\bibinfo {collaboration} {MINOS+}),\ }\bibfield
  {title} {\enquote {\bibinfo {title} {{Search for sterile neutrinos in MINOS
  and MINOS+ using a two-detector fit}},}\ }\href {\doibase
  10.1103/PhysRevLett.122.091803} {\bibfield  {journal} {\bibinfo  {journal}
  {Phys. Rev. Lett.}\ }\textbf {\bibinfo {volume} {122}},\ \bibinfo {pages}
  {091803} (\bibinfo {year} {2019})},\ \Eprint
  {http://arxiv.org/abs/1710.06488} {arXiv:1710.06488 [hep-ex]} \BibitemShut
  {NoStop}%
\bibitem [{\citenamefont {Aartsen}\ \emph {et~al.}(2016)\citenamefont {Aartsen}
  \emph {et~al.}}]{IceCube:2016rnb}%
  \BibitemOpen
  \bibfield  {author} {\bibinfo {author} {\bibfnamefont {M.~G.}\ \bibnamefont
  {Aartsen}} \emph {et~al.} (\bibinfo {collaboration} {IceCube}),\ }\bibfield
  {title} {\enquote {\bibinfo {title} {{Searches for Sterile Neutrinos with the
  IceCube Detector}},}\ }\href {\doibase 10.1103/PhysRevLett.117.071801}
  {\bibfield  {journal} {\bibinfo  {journal} {Phys. Rev. Lett.}\ }\textbf
  {\bibinfo {volume} {117}},\ \bibinfo {pages} {071801} (\bibinfo {year}
  {2016})},\ \Eprint {http://arxiv.org/abs/1605.01990} {arXiv:1605.01990
  [hep-ex]} \BibitemShut {NoStop}%
\bibitem [{\citenamefont {Aartsen}\ \emph
  {et~al.}(2020{\natexlab{a}})\citenamefont {Aartsen} \emph
  {et~al.}}]{IceCube:2020tka}%
  \BibitemOpen
  \bibfield  {author} {\bibinfo {author} {\bibfnamefont {M.~G.}\ \bibnamefont
  {Aartsen}} \emph {et~al.} (\bibinfo {collaboration} {IceCube}),\ }\bibfield
  {title} {\enquote {\bibinfo {title} {{Searching for eV-scale sterile
  neutrinos with eight years of atmospheric neutrinos at the IceCube Neutrino
  Telescope}},}\ }\href {\doibase 10.1103/PhysRevD.102.052009} {\bibfield
  {journal} {\bibinfo  {journal} {Phys. Rev. D}\ }\textbf {\bibinfo {volume}
  {102}},\ \bibinfo {pages} {052009} (\bibinfo {year} {2020}{\natexlab{a}})},\
  \Eprint {http://arxiv.org/abs/2005.12943} {arXiv:2005.12943 [hep-ex]}
  \BibitemShut {NoStop}%
\bibitem [{\citenamefont {Aartsen}\ \emph
  {et~al.}(2020{\natexlab{b}})\citenamefont {Aartsen} \emph
  {et~al.}}]{IceCube:2020phf}%
  \BibitemOpen
  \bibfield  {author} {\bibinfo {author} {\bibfnamefont {M.~G.}\ \bibnamefont
  {Aartsen}} \emph {et~al.} (\bibinfo {collaboration} {IceCube}),\ }\bibfield
  {title} {\enquote {\bibinfo {title} {{eV-Scale Sterile Neutrino Search Using
  Eight Years of Atmospheric Muon Neutrino Data from the IceCube Neutrino
  Observatory}},}\ }\href {\doibase 10.1103/PhysRevLett.125.141801} {\bibfield
  {journal} {\bibinfo  {journal} {Phys. Rev. Lett.}\ }\textbf {\bibinfo
  {volume} {125}},\ \bibinfo {pages} {141801} (\bibinfo {year}
  {2020}{\natexlab{b}})},\ \Eprint {http://arxiv.org/abs/2005.12942}
  {arXiv:2005.12942 [hep-ex]} \BibitemShut {NoStop}%
\bibitem [{\citenamefont {Dentler}\ \emph {et~al.}(2018)\citenamefont
  {Dentler}, \citenamefont {Hern\'andez-Cabezudo}, \citenamefont {Kopp},
  \citenamefont {Machado}, \citenamefont {Maltoni}, \citenamefont
  {Martinez-Soler},\ and\ \citenamefont {Schwetz}}]{Dentler:2018sju}%
  \BibitemOpen
  \bibfield  {author} {\bibinfo {author} {\bibfnamefont {Mona}\ \bibnamefont
  {Dentler}}, \bibinfo {author} {\bibfnamefont {\'Alvaro}\ \bibnamefont
  {Hern\'andez-Cabezudo}}, \bibinfo {author} {\bibfnamefont {Joachim}\
  \bibnamefont {Kopp}}, \bibinfo {author} {\bibfnamefont {Pedro A.~N.}\
  \bibnamefont {Machado}}, \bibinfo {author} {\bibfnamefont {Michele}\
  \bibnamefont {Maltoni}}, \bibinfo {author} {\bibfnamefont {Ivan}\
  \bibnamefont {Martinez-Soler}}, \ and\ \bibinfo {author} {\bibfnamefont
  {Thomas}\ \bibnamefont {Schwetz}},\ }\bibfield  {title} {\enquote {\bibinfo
  {title} {{Updated Global Analysis of Neutrino Oscillations in the Presence of
  eV-Scale Sterile Neutrinos}},}\ }\href {\doibase 10.1007/JHEP08(2018)010}
  {\bibfield  {journal} {\bibinfo  {journal} {JHEP}\ }\textbf {\bibinfo
  {volume} {08}},\ \bibinfo {pages} {010} (\bibinfo {year} {2018})},\ \Eprint
  {http://arxiv.org/abs/1803.10661} {arXiv:1803.10661 [hep-ph]} \BibitemShut
  {NoStop}%
\bibitem [{\citenamefont {Giunti}\ and\ \citenamefont
  {Lasserre}(2019)}]{Giunti:2019aiy}%
  \BibitemOpen
  \bibfield  {author} {\bibinfo {author} {\bibfnamefont {Carlo}\ \bibnamefont
  {Giunti}}\ and\ \bibinfo {author} {\bibfnamefont {T.}~\bibnamefont
  {Lasserre}},\ }\bibfield  {title} {\enquote {\bibinfo {title} {{eV-scale
  Sterile Neutrinos}},}\ }\href {\doibase 10.1146/annurev-nucl-101918-023755}
  {\bibfield  {journal} {\bibinfo  {journal} {Ann. Rev. Nucl. Part. Sci.}\
  }\textbf {\bibinfo {volume} {69}},\ \bibinfo {pages} {163--190} (\bibinfo
  {year} {2019})},\ \Eprint {http://arxiv.org/abs/1901.08330} {arXiv:1901.08330
  [hep-ph]} \BibitemShut {NoStop}%
\bibitem [{\citenamefont {Diaz}\ \emph {et~al.}(2020)\citenamefont {Diaz},
  \citenamefont {Arg\"uelles}, \citenamefont {Collin}, \citenamefont {Conrad},\
  and\ \citenamefont {Shaevitz}}]{Diaz:2019fwt}%
  \BibitemOpen
  \bibfield  {author} {\bibinfo {author} {\bibfnamefont {A.}~\bibnamefont
  {Diaz}}, \bibinfo {author} {\bibfnamefont {C.~A.}\ \bibnamefont
  {Arg\"uelles}}, \bibinfo {author} {\bibfnamefont {G.~H.}\ \bibnamefont
  {Collin}}, \bibinfo {author} {\bibfnamefont {J.~M.}\ \bibnamefont {Conrad}},
  \ and\ \bibinfo {author} {\bibfnamefont {M.~H.}\ \bibnamefont {Shaevitz}},\
  }\bibfield  {title} {\enquote {\bibinfo {title} {{Where Are We With Light
  Sterile Neutrinos?}}}\ }\href {\doibase 10.1016/j.physrep.2020.08.005}
  {\bibfield  {journal} {\bibinfo  {journal} {Phys. Rept.}\ }\textbf {\bibinfo
  {volume} {884}},\ \bibinfo {pages} {1--59} (\bibinfo {year} {2020})},\
  \Eprint {http://arxiv.org/abs/1906.00045} {arXiv:1906.00045 [hep-ex]}
  \BibitemShut {NoStop}%
\bibitem [{\citenamefont {B\"oser}\ \emph {et~al.}(2020)\citenamefont
  {B\"oser}, \citenamefont {Buck}, \citenamefont {Giunti}, \citenamefont
  {Lesgourgues}, \citenamefont {Ludhova}, \citenamefont {Mertens},
  \citenamefont {Schukraft},\ and\ \citenamefont {Wurm}}]{Boser:2019rta}%
  \BibitemOpen
  \bibfield  {author} {\bibinfo {author} {\bibfnamefont {Sebastian}\
  \bibnamefont {B\"oser}}, \bibinfo {author} {\bibfnamefont {Christian}\
  \bibnamefont {Buck}}, \bibinfo {author} {\bibfnamefont {Carlo}\ \bibnamefont
  {Giunti}}, \bibinfo {author} {\bibfnamefont {Julien}\ \bibnamefont
  {Lesgourgues}}, \bibinfo {author} {\bibfnamefont {Livia}\ \bibnamefont
  {Ludhova}}, \bibinfo {author} {\bibfnamefont {Susanne}\ \bibnamefont
  {Mertens}}, \bibinfo {author} {\bibfnamefont {Anne}\ \bibnamefont
  {Schukraft}}, \ and\ \bibinfo {author} {\bibfnamefont {Michael}\ \bibnamefont
  {Wurm}},\ }\bibfield  {title} {\enquote {\bibinfo {title} {{Status of Light
  Sterile Neutrino Searches}},}\ }\href {\doibase 10.1016/j.ppnp.2019.103736}
  {\bibfield  {journal} {\bibinfo  {journal} {Prog. Part. Nucl. Phys.}\
  }\textbf {\bibinfo {volume} {111}},\ \bibinfo {pages} {103736} (\bibinfo
  {year} {2020})},\ \Eprint {http://arxiv.org/abs/1906.01739} {arXiv:1906.01739
  [hep-ex]} \BibitemShut {NoStop}%
\bibitem [{\citenamefont {Eliezer}\ and\ \citenamefont
  {Swift}(1976)}]{Eliezer:1975ja}%
  \BibitemOpen
  \bibfield  {author} {\bibinfo {author} {\bibfnamefont {Shalom}\ \bibnamefont
  {Eliezer}}\ and\ \bibinfo {author} {\bibfnamefont {Arthur~R.}\ \bibnamefont
  {Swift}},\ }\bibfield  {title} {\enquote {\bibinfo {title} {{Experimental
  Consequences of electron Neutrino-Muon-neutrino Mixing in Neutrino Beams}},}\
  }\href {\doibase 10.1016/0550-3213(76)90059-6} {\bibfield  {journal}
  {\bibinfo  {journal} {Nucl. Phys. B}\ }\textbf {\bibinfo {volume} {105}},\
  \bibinfo {pages} {45--51} (\bibinfo {year} {1976})}\BibitemShut {NoStop}%
\bibitem [{\citenamefont {Fritzsch}\ and\ \citenamefont
  {Minkowski}(1976)}]{Fritzsch:1975rz}%
  \BibitemOpen
  \bibfield  {author} {\bibinfo {author} {\bibfnamefont {Harald}\ \bibnamefont
  {Fritzsch}}\ and\ \bibinfo {author} {\bibfnamefont {Peter}\ \bibnamefont
  {Minkowski}},\ }\bibfield  {title} {\enquote {\bibinfo {title} {{Vector-Like
  Weak Currents, Massive Neutrinos, and Neutrino Beam Oscillations}},}\ }\href
  {\doibase 10.1016/0370-2693(76)90051-4} {\bibfield  {journal} {\bibinfo
  {journal} {Phys. Lett. B}\ }\textbf {\bibinfo {volume} {62}},\ \bibinfo
  {pages} {72--76} (\bibinfo {year} {1976})}\BibitemShut {NoStop}%
\bibitem [{\citenamefont {Bilenky}\ and\ \citenamefont
  {Pontecorvo}(1976)}]{Bilenky:1976yj}%
  \BibitemOpen
  \bibfield  {author} {\bibinfo {author} {\bibfnamefont {Samoil~M.}\
  \bibnamefont {Bilenky}}\ and\ \bibinfo {author} {\bibfnamefont
  {B.}~\bibnamefont {Pontecorvo}},\ }\bibfield  {title} {\enquote {\bibinfo
  {title} {{Again on Neutrino Oscillations}},}\ }\href {\doibase
  10.1007/BF02746567} {\bibfield  {journal} {\bibinfo  {journal} {Lett. Nuovo
  Cim.}\ }\textbf {\bibinfo {volume} {17}},\ \bibinfo {pages} {569} (\bibinfo
  {year} {1976})}\BibitemShut {NoStop}%
\bibitem [{\citenamefont {Akhmedov}(2019)}]{Akhmedov:2019iyt}%
  \BibitemOpen
  \bibfield  {author} {\bibinfo {author} {\bibfnamefont {Evgeny}\ \bibnamefont
  {Akhmedov}},\ }\bibfield  {title} {\enquote {\bibinfo {title} {{Quantum
  mechanics aspects and subtleties of neutrino oscillations}},}\ }in\
  \href@noop {} {\emph {\bibinfo {booktitle} {{International Conference on
  History of the Neutrino}: {1930-2018}}}}\ (\bibinfo {year} {2019})\ \Eprint
  {http://arxiv.org/abs/1901.05232} {arXiv:1901.05232 [hep-ph]} \BibitemShut
  {NoStop}%
\bibitem [{\citenamefont {Giunti}(2004)}]{Giunti:2003ax}%
  \BibitemOpen
  \bibfield  {author} {\bibinfo {author} {\bibfnamefont {C.}~\bibnamefont
  {Giunti}},\ }\bibfield  {title} {\enquote {\bibinfo {title} {{Coherence and
  wave packets in neutrino oscillations}},}\ }\href {\doibase
  10.1023/B:FOPL.0000019651.53280.31} {\bibfield  {journal} {\bibinfo
  {journal} {Found. Phys. Lett.}\ }\textbf {\bibinfo {volume} {17}},\ \bibinfo
  {pages} {103--124} (\bibinfo {year} {2004})},\ \Eprint
  {http://arxiv.org/abs/hep-ph/0302026} {arXiv:hep-ph/0302026} \BibitemShut
  {NoStop}%
\bibitem [{\citenamefont {Akhmedov}\ and\ \citenamefont
  {Smirnov}(2009)}]{Akhmedov:2009rb}%
  \BibitemOpen
  \bibfield  {author} {\bibinfo {author} {\bibfnamefont {Evgeny~Kh.}\
  \bibnamefont {Akhmedov}}\ and\ \bibinfo {author} {\bibfnamefont {Alexei~Yu.}\
  \bibnamefont {Smirnov}},\ }\bibfield  {title} {\enquote {\bibinfo {title}
  {{Paradoxes of neutrino oscillations}},}\ }\href {\doibase
  10.1134/S1063778809080122} {\bibfield  {journal} {\bibinfo  {journal} {Phys.
  Atom. Nucl.}\ }\textbf {\bibinfo {volume} {72}},\ \bibinfo {pages}
  {1363--1381} (\bibinfo {year} {2009})},\ \Eprint
  {http://arxiv.org/abs/0905.1903} {arXiv:0905.1903 [hep-ph]} \BibitemShut
  {NoStop}%
\bibitem [{\citenamefont {Nussinov}(1976)}]{Nussinov:1976uw}%
  \BibitemOpen
  \bibfield  {author} {\bibinfo {author} {\bibfnamefont {S.}~\bibnamefont
  {Nussinov}},\ }\bibfield  {title} {\enquote {\bibinfo {title} {{Solar
  Neutrinos and Neutrino Mixing}},}\ }\href {\doibase
  10.1016/0370-2693(76)90648-1} {\bibfield  {journal} {\bibinfo  {journal}
  {Phys. Lett. B}\ }\textbf {\bibinfo {volume} {63}},\ \bibinfo {pages}
  {201--203} (\bibinfo {year} {1976})}\BibitemShut {NoStop}%
\bibitem [{\citenamefont {Kayser}(1981)}]{Kayser:1981ye}%
  \BibitemOpen
  \bibfield  {author} {\bibinfo {author} {\bibfnamefont {Boris}\ \bibnamefont
  {Kayser}},\ }\bibfield  {title} {\enquote {\bibinfo {title} {{On the Quantum
  Mechanics of Neutrino Oscillation}},}\ }\href {\doibase
  10.1103/PhysRevD.24.110} {\bibfield  {journal} {\bibinfo  {journal} {Phys.
  Rev. D}\ }\textbf {\bibinfo {volume} {24}},\ \bibinfo {pages} {110} (\bibinfo
  {year} {1981})}\BibitemShut {NoStop}%
\bibitem [{\citenamefont {Kiers}\ \emph {et~al.}(1996)\citenamefont {Kiers},
  \citenamefont {Nussinov},\ and\ \citenamefont {Weiss}}]{Kiers:1995zj}%
  \BibitemOpen
  \bibfield  {author} {\bibinfo {author} {\bibfnamefont {Ken}\ \bibnamefont
  {Kiers}}, \bibinfo {author} {\bibfnamefont {Shmuel}\ \bibnamefont
  {Nussinov}}, \ and\ \bibinfo {author} {\bibfnamefont {Nathan}\ \bibnamefont
  {Weiss}},\ }\bibfield  {title} {\enquote {\bibinfo {title} {{Coherence
  effects in neutrino oscillations}},}\ }\href {\doibase
  10.1103/PhysRevD.53.537} {\bibfield  {journal} {\bibinfo  {journal} {Phys.
  Rev. D}\ }\textbf {\bibinfo {volume} {53}},\ \bibinfo {pages} {537--547}
  (\bibinfo {year} {1996})},\ \Eprint {http://arxiv.org/abs/hep-ph/9506271}
  {arXiv:hep-ph/9506271} \BibitemShut {NoStop}%
\bibitem [{\citenamefont {Beuthe}(2003)}]{Beuthe:2001rc}%
  \BibitemOpen
  \bibfield  {author} {\bibinfo {author} {\bibfnamefont {Mikael}\ \bibnamefont
  {Beuthe}},\ }\bibfield  {title} {\enquote {\bibinfo {title} {{Oscillations of
  neutrinos and mesons in quantum field theory}},}\ }\href {\doibase
  10.1016/S0370-1573(02)00538-0} {\bibfield  {journal} {\bibinfo  {journal}
  {Phys. Rept.}\ }\textbf {\bibinfo {volume} {375}},\ \bibinfo {pages}
  {105--218} (\bibinfo {year} {2003})},\ \Eprint
  {http://arxiv.org/abs/hep-ph/0109119} {arXiv:hep-ph/0109119} \BibitemShut
  {NoStop}%
\bibitem [{\citenamefont {Akhmedov}\ \emph {et~al.}(2012)\citenamefont
  {Akhmedov}, \citenamefont {Hernandez},\ and\ \citenamefont
  {Smirnov}}]{Akhmedov:2012uu}%
  \BibitemOpen
  \bibfield  {author} {\bibinfo {author} {\bibfnamefont {Evgeny}\ \bibnamefont
  {Akhmedov}}, \bibinfo {author} {\bibfnamefont {Daniel}\ \bibnamefont
  {Hernandez}}, \ and\ \bibinfo {author} {\bibfnamefont {Alexei}\ \bibnamefont
  {Smirnov}},\ }\bibfield  {title} {\enquote {\bibinfo {title} {{Neutrino
  production coherence and oscillation experiments}},}\ }\href {\doibase
  10.1007/JHEP04(2012)052} {\bibfield  {journal} {\bibinfo  {journal} {JHEP}\
  }\textbf {\bibinfo {volume} {04}},\ \bibinfo {pages} {052} (\bibinfo {year}
  {2012})},\ \Eprint {http://arxiv.org/abs/1201.4128} {arXiv:1201.4128
  [hep-ph]} \BibitemShut {NoStop}%
\bibitem [{\citenamefont {Akhmedov}\ \emph {et~al.}(2017)\citenamefont
  {Akhmedov}, \citenamefont {Kopp},\ and\ \citenamefont
  {Lindner}}]{Akhmedov:2017mcc}%
  \BibitemOpen
  \bibfield  {author} {\bibinfo {author} {\bibfnamefont {Evgeny}\ \bibnamefont
  {Akhmedov}}, \bibinfo {author} {\bibfnamefont {Joachim}\ \bibnamefont
  {Kopp}}, \ and\ \bibinfo {author} {\bibfnamefont {Manfred}\ \bibnamefont
  {Lindner}},\ }\bibfield  {title} {\enquote {\bibinfo {title} {{Collective
  neutrino oscillations and neutrino wave packets}},}\ }\href {\doibase
  10.1088/1475-7516/2017/09/017} {\bibfield  {journal} {\bibinfo  {journal}
  {JCAP}\ }\textbf {\bibinfo {volume} {09}},\ \bibinfo {pages} {017} (\bibinfo
  {year} {2017})},\ \Eprint {http://arxiv.org/abs/1702.08338} {arXiv:1702.08338
  [hep-ph]} \BibitemShut {NoStop}%
\bibitem [{\citenamefont {Giunti}\ and\ \citenamefont
  {Kim}(1998)}]{Giunti:1997wq}%
  \BibitemOpen
  \bibfield  {author} {\bibinfo {author} {\bibfnamefont {C.}~\bibnamefont
  {Giunti}}\ and\ \bibinfo {author} {\bibfnamefont {C.~W.}\ \bibnamefont
  {Kim}},\ }\bibfield  {title} {\enquote {\bibinfo {title} {{Coherence of
  neutrino oscillations in the wave packet approach}},}\ }\href {\doibase
  10.1103/PhysRevD.58.017301} {\bibfield  {journal} {\bibinfo  {journal} {Phys.
  Rev. D}\ }\textbf {\bibinfo {volume} {58}},\ \bibinfo {pages} {017301}
  (\bibinfo {year} {1998})},\ \Eprint {http://arxiv.org/abs/hep-ph/9711363}
  {arXiv:hep-ph/9711363} \BibitemShut {NoStop}%
\bibitem [{\citenamefont {Giunti}\ and\ \citenamefont
  {Kim}(2007)}]{Giunti:2007ry}%
  \BibitemOpen
  \bibfield  {author} {\bibinfo {author} {\bibfnamefont {Carlo}\ \bibnamefont
  {Giunti}}\ and\ \bibinfo {author} {\bibfnamefont {Chung~W.}\ \bibnamefont
  {Kim}},\ }\href@noop {} {\emph {\bibinfo {title} {{Fundamentals of Neutrino
  Physics and Astrophysics}}}}\ (\bibinfo {year} {2007})\BibitemShut {NoStop}%
\bibitem [{\citenamefont {Bernardini}\ and\ \citenamefont
  {De~Leo}(2004)}]{Bernardini:2004sw}%
  \BibitemOpen
  \bibfield  {author} {\bibinfo {author} {\bibfnamefont {Alex~E.}\ \bibnamefont
  {Bernardini}}\ and\ \bibinfo {author} {\bibfnamefont {Stefano}\ \bibnamefont
  {De~Leo}},\ }\bibfield  {title} {\enquote {\bibinfo {title} {{An Analytic
  approach to the wave packet formalism in oscillation phenomena}},}\ }\href
  {\doibase 10.1103/PhysRevD.70.053010} {\bibfield  {journal} {\bibinfo
  {journal} {Phys. Rev. D}\ }\textbf {\bibinfo {volume} {70}},\ \bibinfo
  {pages} {053010} (\bibinfo {year} {2004})},\ \Eprint
  {http://arxiv.org/abs/hep-ph/0411134} {arXiv:hep-ph/0411134} \BibitemShut
  {NoStop}%
\bibitem [{\citenamefont {Naumov}\ and\ \citenamefont
  {Naumov}(2010)}]{Naumov:2010um}%
  \BibitemOpen
  \bibfield  {author} {\bibinfo {author} {\bibfnamefont {D.~V.}\ \bibnamefont
  {Naumov}}\ and\ \bibinfo {author} {\bibfnamefont {V.~A.}\ \bibnamefont
  {Naumov}},\ }\bibfield  {title} {\enquote {\bibinfo {title} {{A Diagrammatic
  treatment of neutrino oscillations}},}\ }\href {\doibase
  10.1088/0954-3899/37/10/105014} {\bibfield  {journal} {\bibinfo  {journal}
  {J. Phys. G}\ }\textbf {\bibinfo {volume} {37}},\ \bibinfo {pages} {105014}
  (\bibinfo {year} {2010})},\ \Eprint {http://arxiv.org/abs/1008.0306}
  {arXiv:1008.0306 [hep-ph]} \BibitemShut {NoStop}%
\bibitem [{\citenamefont {Jones}(2015)}]{Jones:2014sfa}%
  \BibitemOpen
  \bibfield  {author} {\bibinfo {author} {\bibfnamefont {B.~J.~P.}\
  \bibnamefont {Jones}},\ }\bibfield  {title} {\enquote {\bibinfo {title}
  {{Dynamical pion collapse and the coherence of conventional neutrino
  beams}},}\ }\href {\doibase 10.1103/PhysRevD.91.053002} {\bibfield  {journal}
  {\bibinfo  {journal} {Phys. Rev. D}\ }\textbf {\bibinfo {volume} {91}},\
  \bibinfo {pages} {053002} (\bibinfo {year} {2015})},\ \Eprint
  {http://arxiv.org/abs/1412.2264} {arXiv:1412.2264 [hep-ph]} \BibitemShut
  {NoStop}%
\bibitem [{\citenamefont {An}\ \emph {et~al.}(2017{\natexlab{b}})\citenamefont
  {An} \emph {et~al.}}]{DayaBay:2016ouy}%
  \BibitemOpen
  \bibfield  {author} {\bibinfo {author} {\bibfnamefont {Feng~Peng}\
  \bibnamefont {An}} \emph {et~al.} (\bibinfo {collaboration} {Daya Bay}),\
  }\bibfield  {title} {\enquote {\bibinfo {title} {{Study of the wave packet
  treatment of neutrino oscillation at Daya Bay}},}\ }\href {\doibase
  10.1140/epjc/s10052-017-4970-y} {\bibfield  {journal} {\bibinfo  {journal}
  {Eur. Phys. J. C}\ }\textbf {\bibinfo {volume} {77}},\ \bibinfo {pages} {606}
  (\bibinfo {year} {2017}{\natexlab{b}})},\ \Eprint
  {http://arxiv.org/abs/1608.01661} {arXiv:1608.01661 [hep-ex]} \BibitemShut
  {NoStop}%
\bibitem [{\citenamefont {Akhmedov}\ and\ \citenamefont
  {Smirnov}(2022)}]{Akhmedov:2022bjs}%
  \BibitemOpen
  \bibfield  {author} {\bibinfo {author} {\bibfnamefont {Evgeny}\ \bibnamefont
  {Akhmedov}}\ and\ \bibinfo {author} {\bibfnamefont {Alexei~Y.}\ \bibnamefont
  {Smirnov}},\ }\bibfield  {title} {\enquote {\bibinfo {title} {{Damping of
  neutrino oscillations, decoherence and the lengths of neutrino wave
  packets}},}\ }\href@noop {} {\  (\bibinfo {year} {2022})},\ \Eprint
  {http://arxiv.org/abs/2208.03736} {arXiv:2208.03736 [hep-ph]} \BibitemShut
  {NoStop}%
\bibitem [{\citenamefont {Jones}(2022)}]{Jones:2022cvh}%
  \BibitemOpen
  \bibfield  {author} {\bibinfo {author} {\bibfnamefont {B.~J.~P.}\
  \bibnamefont {Jones}},\ }\bibfield  {title} {\enquote {\bibinfo {title}
  {{Comment on ''Damping of neutrino oscillations, decoherence and the lengths
  of neutrino wave packets''}},}\ }\href@noop {} {\  (\bibinfo {year}
  {2022})},\ \Eprint {http://arxiv.org/abs/2209.00561} {arXiv:2209.00561
  [hep-ph]} \BibitemShut {NoStop}%
\bibitem [{\citenamefont {Jones}\ \emph {et~al.}(2022)\citenamefont {Jones},
  \citenamefont {Marzec},\ and\ \citenamefont {Spitz}}]{Jones:2022hme}%
  \BibitemOpen
  \bibfield  {author} {\bibinfo {author} {\bibfnamefont {B.~J.~P.}\
  \bibnamefont {Jones}}, \bibinfo {author} {\bibfnamefont {E.}~\bibnamefont
  {Marzec}}, \ and\ \bibinfo {author} {\bibfnamefont {J.}~\bibnamefont
  {Spitz}},\ }\bibfield  {title} {\enquote {\bibinfo {title} {{The Width of a
  Beta-decay-induced Antineutrino Wavepacket}},}\ }\href@noop {} {\  (\bibinfo
  {year} {2022})},\ \Eprint {http://arxiv.org/abs/2211.00026} {arXiv:2211.00026
  [hep-ph]} \BibitemShut {NoStop}%
\bibitem [{\citenamefont {de~Gouv\^ea}\ \emph {et~al.}(2021)\citenamefont
  {de~Gouv\^ea}, \citenamefont {De~Romeri},\ and\ \citenamefont
  {Ternes}}]{deGouvea:2021uvg}%
  \BibitemOpen
  \bibfield  {author} {\bibinfo {author} {\bibfnamefont {Andr\'e}\ \bibnamefont
  {de~Gouv\^ea}}, \bibinfo {author} {\bibfnamefont {Valentina}\ \bibnamefont
  {De~Romeri}}, \ and\ \bibinfo {author} {\bibfnamefont {Christoph~A.}\
  \bibnamefont {Ternes}},\ }\bibfield  {title} {\enquote {\bibinfo {title}
  {{Combined analysis of neutrino decoherence at reactor experiments}},}\
  }\href {\doibase 10.1007/JHEP06(2021)042} {\bibfield  {journal} {\bibinfo
  {journal} {JHEP}\ }\textbf {\bibinfo {volume} {06}},\ \bibinfo {pages} {042}
  (\bibinfo {year} {2021})},\ \Eprint {http://arxiv.org/abs/2104.05806}
  {arXiv:2104.05806 [hep-ph]} \BibitemShut {NoStop}%
\bibitem [{\citenamefont {Stodolsky}(1998)}]{Stodolsky:1998tc}%
  \BibitemOpen
  \bibfield  {author} {\bibinfo {author} {\bibfnamefont {Leo}\ \bibnamefont
  {Stodolsky}},\ }\bibfield  {title} {\enquote {\bibinfo {title} {{The
  Unnecessary wave packet}},}\ }\href {\doibase 10.1103/PhysRevD.58.036006}
  {\bibfield  {journal} {\bibinfo  {journal} {Phys. Rev. D}\ }\textbf {\bibinfo
  {volume} {58}},\ \bibinfo {pages} {036006} (\bibinfo {year} {1998})},\
  \Eprint {http://arxiv.org/abs/hep-ph/9802387} {arXiv:hep-ph/9802387}
  \BibitemShut {NoStop}%
\bibitem [{\citenamefont {Giunti}\ \emph {et~al.}(1991)\citenamefont {Giunti},
  \citenamefont {Kim},\ and\ \citenamefont {Lee}}]{Giunti:1991ca}%
  \BibitemOpen
  \bibfield  {author} {\bibinfo {author} {\bibfnamefont {C.}~\bibnamefont
  {Giunti}}, \bibinfo {author} {\bibfnamefont {C.~W.}\ \bibnamefont {Kim}}, \
  and\ \bibinfo {author} {\bibfnamefont {U.~W.}\ \bibnamefont {Lee}},\
  }\bibfield  {title} {\enquote {\bibinfo {title} {{When do neutrinos really
  oscillate?: Quantum mechanics of neutrino oscillations}},}\ }\href {\doibase
  10.1103/PhysRevD.44.3635} {\bibfield  {journal} {\bibinfo  {journal} {Phys.
  Rev. D}\ }\textbf {\bibinfo {volume} {44}},\ \bibinfo {pages} {3635--3640}
  (\bibinfo {year} {1991})}\BibitemShut {NoStop}%
\bibitem [{\citenamefont {Akhmedov}(2017)}]{Akhmedov:2017xxm}%
  \BibitemOpen
  \bibfield  {author} {\bibinfo {author} {\bibfnamefont {Evgeny}\ \bibnamefont
  {Akhmedov}},\ }\bibfield  {title} {\enquote {\bibinfo {title} {{Do
  non-relativistic neutrinos oscillate?}}}\ }\href {\doibase
  10.1007/JHEP07(2017)070} {\bibfield  {journal} {\bibinfo  {journal} {JHEP}\
  }\textbf {\bibinfo {volume} {07}},\ \bibinfo {pages} {070} (\bibinfo {year}
  {2017})},\ \Eprint {http://arxiv.org/abs/1703.08169} {arXiv:1703.08169
  [hep-ph]} \BibitemShut {NoStop}%
\bibitem [{\citenamefont {Dentler}\ \emph {et~al.}(2017)\citenamefont
  {Dentler}, \citenamefont {Hern\'andez-Cabezudo}, \citenamefont {Kopp},
  \citenamefont {Maltoni},\ and\ \citenamefont {Schwetz}}]{Dentler:2017tkw}%
  \BibitemOpen
  \bibfield  {author} {\bibinfo {author} {\bibfnamefont {Mona}\ \bibnamefont
  {Dentler}}, \bibinfo {author} {\bibfnamefont {\'Alvaro}\ \bibnamefont
  {Hern\'andez-Cabezudo}}, \bibinfo {author} {\bibfnamefont {Joachim}\
  \bibnamefont {Kopp}}, \bibinfo {author} {\bibfnamefont {Michele}\
  \bibnamefont {Maltoni}}, \ and\ \bibinfo {author} {\bibfnamefont {Thomas}\
  \bibnamefont {Schwetz}},\ }\bibfield  {title} {\enquote {\bibinfo {title}
  {{Sterile neutrinos or flux uncertainties? \textemdash{} Status of the
  reactor anti-neutrino anomaly}},}\ }\href {\doibase 10.1007/JHEP11(2017)099}
  {\bibfield  {journal} {\bibinfo  {journal} {JHEP}\ }\textbf {\bibinfo
  {volume} {11}},\ \bibinfo {pages} {099} (\bibinfo {year} {2017})},\ \Eprint
  {http://arxiv.org/abs/1709.04294} {arXiv:1709.04294 [hep-ph]} \BibitemShut
  {NoStop}%
\bibitem [{\citenamefont {Huber}(2011)}]{Huber:2011wv}%
  \BibitemOpen
  \bibfield  {author} {\bibinfo {author} {\bibfnamefont {Patrick}\ \bibnamefont
  {Huber}},\ }\bibfield  {title} {\enquote {\bibinfo {title} {{On the
  determination of anti-neutrino spectra from nuclear reactors}},}\ }\href
  {\doibase 10.1103/PhysRevC.85.029901} {\bibfield  {journal} {\bibinfo
  {journal} {Phys. Rev. C}\ }\textbf {\bibinfo {volume} {84}},\ \bibinfo
  {pages} {024617} (\bibinfo {year} {2011})},\ \bibinfo {note} {[Erratum:
  Phys.Rev.C 85, 029901 (2012)]},\ \Eprint {http://arxiv.org/abs/1106.0687}
  {arXiv:1106.0687 [hep-ph]} \BibitemShut {NoStop}%
\bibitem [{\citenamefont {Mueller}\ \emph {et~al.}(2011)\citenamefont {Mueller}
  \emph {et~al.}}]{Mueller:2011nm}%
  \BibitemOpen
  \bibfield  {author} {\bibinfo {author} {\bibfnamefont {Th.~A.}\ \bibnamefont
  {Mueller}} \emph {et~al.},\ }\bibfield  {title} {\enquote {\bibinfo {title}
  {{Improved Predictions of Reactor Antineutrino Spectra}},}\ }\href {\doibase
  10.1103/PhysRevC.83.054615} {\bibfield  {journal} {\bibinfo  {journal} {Phys.
  Rev. C}\ }\textbf {\bibinfo {volume} {83}},\ \bibinfo {pages} {054615}
  (\bibinfo {year} {2011})},\ \Eprint {http://arxiv.org/abs/1101.2663}
  {arXiv:1101.2663 [hep-ex]} \BibitemShut {NoStop}%
\bibitem [{\citenamefont {Huber}(2016)}]{Huber2016}%
  \BibitemOpen
  \bibfield  {author} {\bibinfo {author} {\bibfnamefont {Patrick}\ \bibnamefont
  {Huber}},\ }\bibfield  {title} {\enquote {\bibinfo {title} {The 5 mev bump -
  a nuclear whodunit mystery},}\ }\href {\doibase
  10.1103/PhysRevLett.118.042502} {\bibfield  {journal} {\bibinfo  {journal}
  {Phys. Rev. Lett. 118, 042502 (2017)}\ } (\bibinfo {year} {2016}),\
  10.1103/PhysRevLett.118.042502},\ \Eprint {http://arxiv.org/abs/1609.03910}
  {arXiv:1609.03910 [hep-ph]} \BibitemShut {NoStop}%
\bibitem [{\citenamefont {Adey}\ \emph {et~al.}(2019)\citenamefont {Adey} \emph
  {et~al.}}]{DayaBay:2018heb}%
  \BibitemOpen
  \bibfield  {author} {\bibinfo {author} {\bibfnamefont {D.}~\bibnamefont
  {Adey}} \emph {et~al.} (\bibinfo {collaboration} {Daya Bay}),\ }\bibfield
  {title} {\enquote {\bibinfo {title} {{Improved Measurement of the Reactor
  Antineutrino Flux at Daya Bay}},}\ }\href {\doibase
  10.1103/PhysRevD.100.052004} {\bibfield  {journal} {\bibinfo  {journal}
  {Phys. Rev. D}\ }\textbf {\bibinfo {volume} {100}},\ \bibinfo {pages}
  {052004} (\bibinfo {year} {2019})},\ \Eprint
  {http://arxiv.org/abs/1808.10836} {arXiv:1808.10836 [hep-ex]} \BibitemShut
  {NoStop}%
\bibitem [{\citenamefont {Giunti}(2002)}]{Giunti:2002xg}%
  \BibitemOpen
  \bibfield  {author} {\bibinfo {author} {\bibfnamefont {C.}~\bibnamefont
  {Giunti}},\ }\bibfield  {title} {\enquote {\bibinfo {title} {{Neutrino wave
  packets in quantum field theory}},}\ }\href {\doibase
  10.1088/1126-6708/2002/11/017} {\bibfield  {journal} {\bibinfo  {journal}
  {JHEP}\ }\textbf {\bibinfo {volume} {11}},\ \bibinfo {pages} {017} (\bibinfo
  {year} {2002})},\ \Eprint {http://arxiv.org/abs/hep-ph/0205014}
  {arXiv:hep-ph/0205014} \BibitemShut {NoStop}%
\bibitem [{\citenamefont {Akhmedov}\ and\ \citenamefont
  {Kopp}(2010)}]{Akhmedov:2010ms}%
  \BibitemOpen
  \bibfield  {author} {\bibinfo {author} {\bibfnamefont {Evgeny~Kh.}\
  \bibnamefont {Akhmedov}}\ and\ \bibinfo {author} {\bibfnamefont {Joachim}\
  \bibnamefont {Kopp}},\ }\bibfield  {title} {\enquote {\bibinfo {title}
  {{Neutrino Oscillations: Quantum Mechanics vs. Quantum Field Theory}},}\
  }\href {\doibase 10.1007/JHEP04(2010)008} {\bibfield  {journal} {\bibinfo
  {journal} {JHEP}\ }\textbf {\bibinfo {volume} {04}},\ \bibinfo {pages} {008}
  (\bibinfo {year} {2010})},\ \bibinfo {note} {[Erratum: JHEP 10, 052
  (2013)]},\ \Eprint {http://arxiv.org/abs/1001.4815} {arXiv:1001.4815
  [hep-ph]} \BibitemShut {NoStop}%
\bibitem [{\citenamefont {Torres}\ \emph {et~al.}(2020)\citenamefont {Torres},
  \citenamefont {Perche}, \citenamefont {Landulfo},\ and\ \citenamefont
  {Matsas}}]{Torres:2020gzm}%
  \BibitemOpen
  \bibfield  {author} {\bibinfo {author} {\bibfnamefont {Bruno de S.~L.}\
  \bibnamefont {Torres}}, \bibinfo {author} {\bibfnamefont {T.~Rick}\
  \bibnamefont {Perche}}, \bibinfo {author} {\bibfnamefont {Andr\'e G.~S.}\
  \bibnamefont {Landulfo}}, \ and\ \bibinfo {author} {\bibfnamefont {George
  E.~A.}\ \bibnamefont {Matsas}},\ }\bibfield  {title} {\enquote {\bibinfo
  {title} {{Neutrino flavor oscillations without flavor states}},}\ }\href
  {\doibase 10.1103/PhysRevD.102.093003} {\bibfield  {journal} {\bibinfo
  {journal} {Phys. Rev. D}\ }\textbf {\bibinfo {volume} {102}},\ \bibinfo
  {pages} {093003} (\bibinfo {year} {2020})},\ \Eprint
  {http://arxiv.org/abs/2009.10165} {arXiv:2009.10165 [hep-ph]} \BibitemShut
  {NoStop}%
\bibitem [{\citenamefont {Oralbaev}\ \emph {et~al.}(2016)\citenamefont
  {Oralbaev}, \citenamefont {Skorokhvatov},\ and\ \citenamefont
  {Titov}}]{Oralbaev_2016}%
  \BibitemOpen
  \bibfield  {author} {\bibinfo {author} {\bibfnamefont {A}~\bibnamefont
  {Oralbaev}}, \bibinfo {author} {\bibfnamefont {M}~\bibnamefont
  {Skorokhvatov}}, \ and\ \bibinfo {author} {\bibfnamefont {O}~\bibnamefont
  {Titov}},\ }\bibfield  {title} {\enquote {\bibinfo {title} {The inverse beta
  decay: a study of cross section},}\ }\href {\doibase
  10.1088/1742-6596/675/1/012003} {\bibfield  {journal} {\bibinfo  {journal}
  {Journal of Physics: Conference Series}\ }\textbf {\bibinfo {volume} {675}},\
  \bibinfo {pages} {012003} (\bibinfo {year} {2016})}\BibitemShut {NoStop}%
\bibitem [{\citenamefont {Esteban}\ \emph {et~al.}(2020)\citenamefont
  {Esteban}, \citenamefont {Gonzalez-Garcia}, \citenamefont {Maltoni},
  \citenamefont {Schwetz},\ and\ \citenamefont {Zhou}}]{Esteban2020}%
  \BibitemOpen
  \bibfield  {author} {\bibinfo {author} {\bibfnamefont {Ivan}\ \bibnamefont
  {Esteban}}, \bibinfo {author} {\bibfnamefont {M.C.}\ \bibnamefont
  {Gonzalez-Garcia}}, \bibinfo {author} {\bibfnamefont {Michele}\ \bibnamefont
  {Maltoni}}, \bibinfo {author} {\bibfnamefont {Thomas}\ \bibnamefont
  {Schwetz}}, \ and\ \bibinfo {author} {\bibfnamefont {Albert}\ \bibnamefont
  {Zhou}},\ }\bibfield  {title} {\enquote {\bibinfo {title} {The fate of hints:
  updated global analysis of three-flavor neutrino oscillations},}\ }\href
  {\doibase 10.1007/jhep09(2020)178} {\ \textbf {\bibinfo {volume} {2020}}
  (\bibinfo {year} {2020}),\ 10.1007/jhep09(2020)178},\ \Eprint
  {http://arxiv.org/abs/http://arxiv.org/abs/2007.14792}
  {arXiv:http://arxiv.org/abs/2007.14792 [hep-ph]} \BibitemShut {NoStop}%
\bibitem [{\citenamefont {Kim}(2017)}]{NEOS:2017}%
  \BibitemOpen
  \bibfield  {author} {\bibinfo {author} {\bibfnamefont {Hyunsoo}\ \bibnamefont
  {Kim}} (\bibinfo {collaboration} {NEOS}),\ }\bibfield  {title} {\enquote
  {\bibinfo {title} {{Search for Sterile Neutrino at NEOS Experiment}},}\
  }\href@noop {} {\bibfield  {journal} {\bibinfo  {journal} {13th Recontres du
  Vietnam}\ } (\bibinfo {year} {2017})}\BibitemShut {NoStop}%
\bibitem [{\citenamefont {Yoon}(2021)}]{RENO:2021}%
  \BibitemOpen
  \bibfield  {author} {\bibinfo {author} {\bibfnamefont {Seok-Gyeong}\
  \bibnamefont {Yoon}} (\bibinfo {collaboration} {RENO}),\ }\bibfield  {title}
  {\enquote {\bibinfo {title} {{Search for sterile neutrinos at RENO}},}\
  }\href@noop {} {\bibfield  {journal} {\bibinfo  {journal} {XIX International
  Workshop on Neutrino Telescopes}\ } (\bibinfo {year} {2021})}\BibitemShut
  {NoStop}%
\end{thebibliography}%

\pagebreak
\clearpage


\onecolumngrid
\appendix

\ifx \standalonesupplemental\undefined
\setcounter{page}{1}
\setcounter{figure}{0}
\setcounter{table}{0}
\setcounter{equation}{0}
\fi

\renewcommand{\thepage}{Supplemental Material-- S\arabic{page}}
\renewcommand{\figurename}{SUPPL. FIG.}
\renewcommand{\tablename}{SUPPL. TABLE}

\renewcommand{\theequation}{A\arabic{equation}}
\clearpage

\begin{center}
\textbf{\large Supplemental Material}
\end{center}

\section*{Results for individual experiments}

\begin{figure*}[ht!]
    \centering
    \includegraphics[width = 0.45\linewidth]{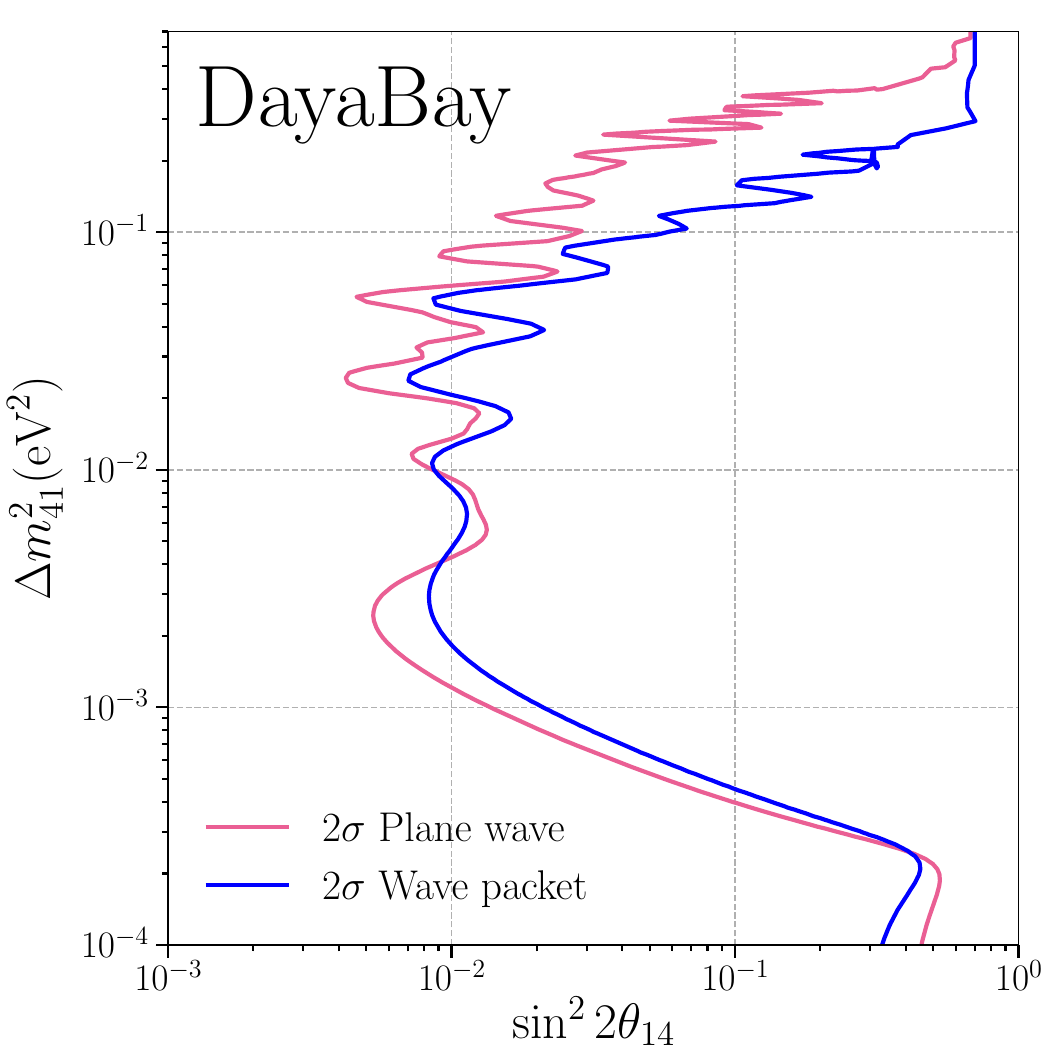}
    \includegraphics[width = 0.45\linewidth]{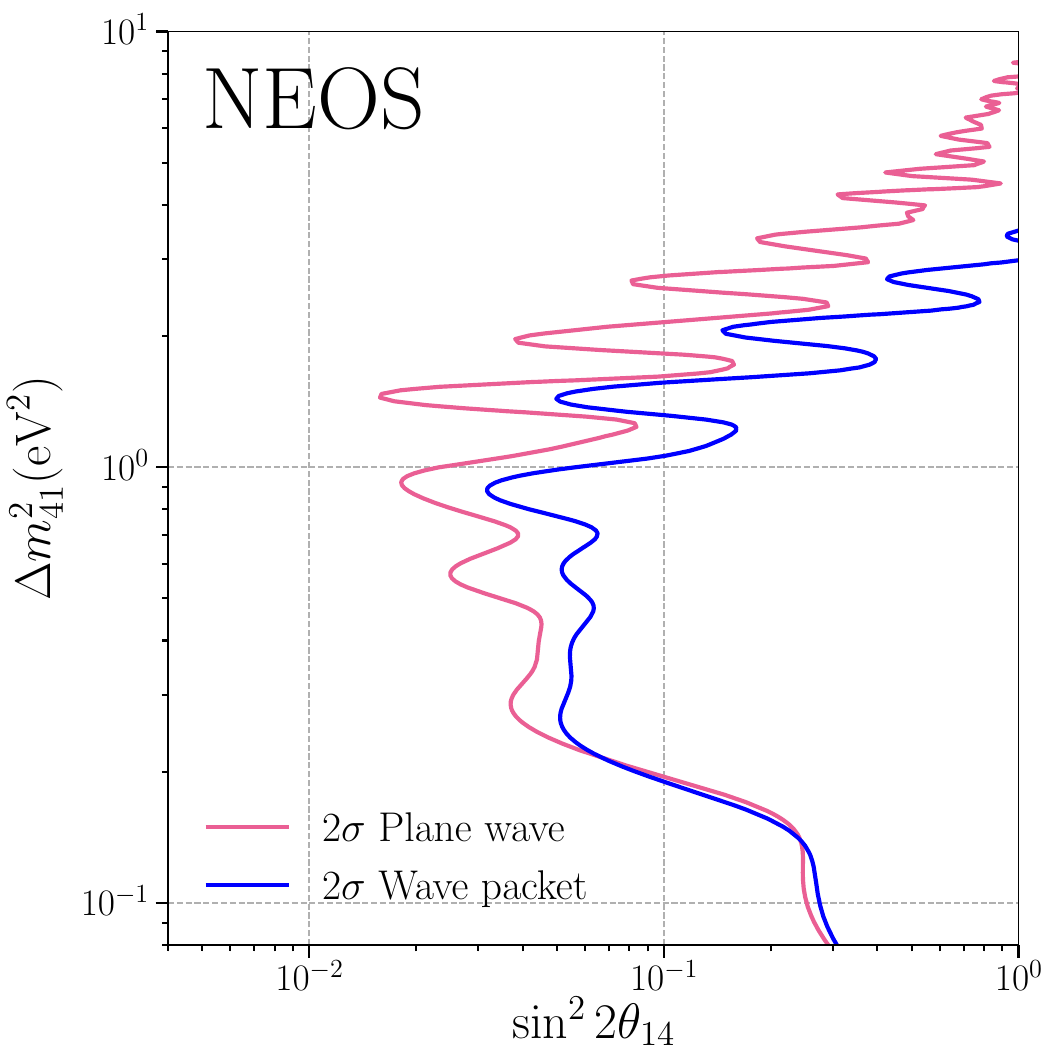}\vspace{6pt}
    
    \includegraphics[width = 0.45\linewidth]{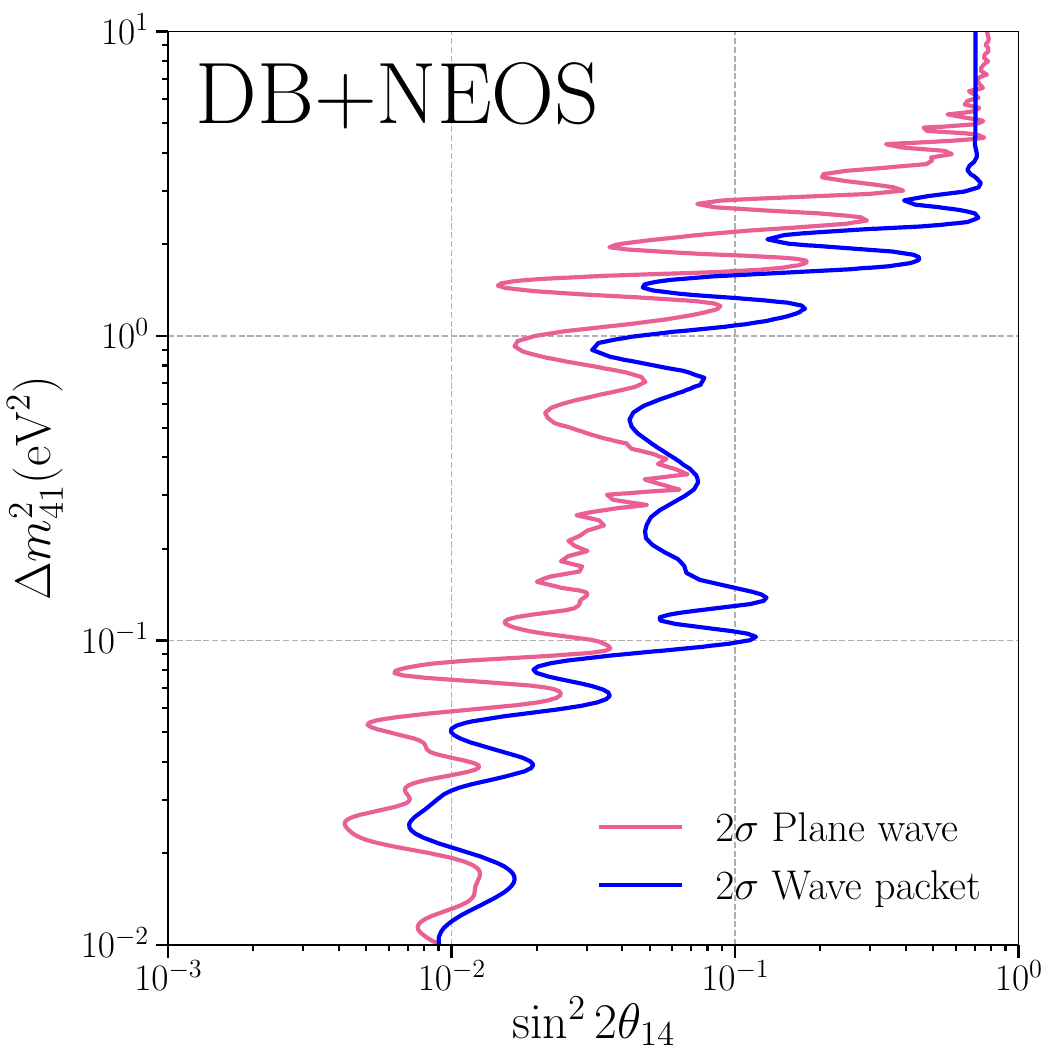}
    \includegraphics[width = 0.45\linewidth]{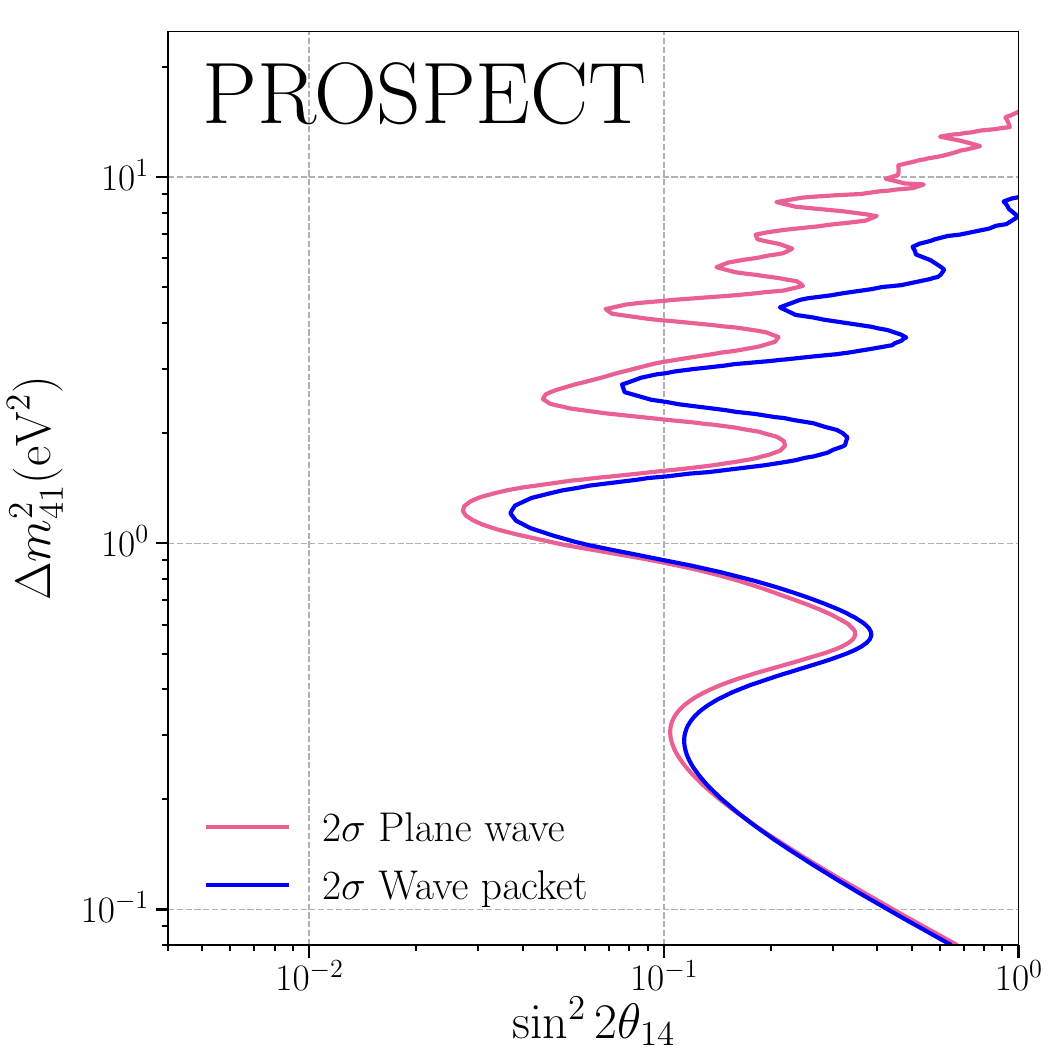}
    \caption{\textbf{\textit{Effect of finite wave packet size on different electron neutrino disappearance experiments and their combination.}} The solid pink and solid blue contours bound the exclusion region at two sigma for the PW and WP formalisms, respectively. All contours are obtained using $\sigma_x = 2.1\times 10^{-4}${\rm nm}~\cite{deGouvea:2021uvg}, and are drawn with respect to the null hypothesis.}
    \label{fig:globalfit}
\end{figure*}

\clearpage

\section*{Results for different wave packet sizes}

\begin{figure*}[ht!]
    \centering
    \includegraphics[width = 0.8\linewidth]{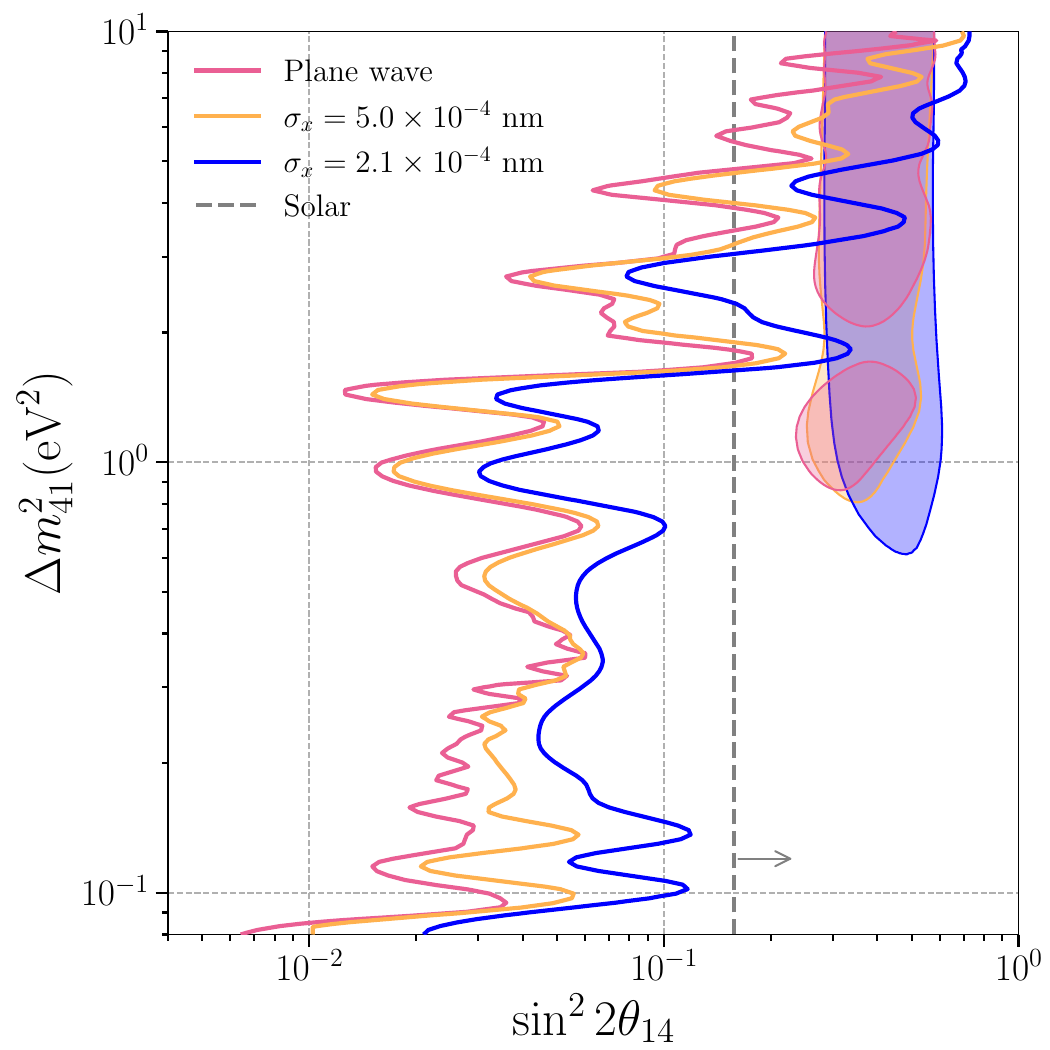}
    \caption{\textbf{\textit{Effect of different finite wave packet sizes on the combination of the experiments considered in this work.}} While the solid pink contour bounds the exclusion region at two sigma in the PW formalism, the solid yellow and solid blue contours are computed in the WP formalism with $\sigma_x = 5.0\times 10^{-4}${\rm nm} and $\sigma_x = 2.1\times 10^{-4}${\rm nm}~\cite{deGouvea:2021uvg}, respectively. A contour with wave packet width $\sigma_x\sim 2\times 10^{-3}${\rm nm} is essentially indistinguishable from the PW contour. All exclusion contours are drawn with respect to the null hypothesis. Furthermore, the preferred region at two sigma for the BEST experiment is shaded for the PW approximation (pink) and for the WP packet formalism, yellow for $\sigma_x = 5.0\times 10^{-4}${\rm nm} and blue for $\sigma_x = 2.1\times 10^{-4}${\rm nm}~\cite{deGouvea:2021uvg}. Finally, a gray dashed line marks the 2 sigma bounds from solar neutrino experiments~\cite{Berryman:2021yan,Goldhagen:2021kxe}.}
    \label{fig:larger-sigmax}
\end{figure*}

\clearpage

\section*{Details of the wave packet formalism}
Let $\nu_\alpha$ be a neutrino of flavor $\alpha$, produced through some weak interaction. In the usual PW approximation, its state has well-defined energy and evolves in time according to~\cite{Giunti:1997wq,Nussinov:1976uw,Kayser:1981ye,Kiers:1995zj,Beuthe:2001rc,Akhmedov:2012uu,Akhmedov:2017mcc}
\begin{equation}
    \ket{\nu_\alpha(\vec{x},t)} = \sum_{i=1}^n U_{\alpha i}^* \ket{\nu_i(t)} = \sum_{i=1}^n U_{\alpha i}^* e^{-iE_i(p)t} \ket{\nu_i(0)},
\end{equation}with $U_{\alpha i}$ being the neutrino mixing matrix, $n$ the total number of neutrino mass eigenstates $\nu_i$ and $E_i(p) = \sqrt{p^2+m_i^2}$ the relativistic expression for the energy. 
This expression assumes that all mass eigenstates have the same momentum $p$, which is obviously in contradiction to the kinematics of any neutrino production process. Nonetheless, this derivation of the oscillation formula leads to the correct result in the regime where the effect of the WP width is negligible~\cite{Akhmedov:2009rb}. 

On the contrary, in the WP formalism the states $\ket{\nu_\alpha(t)}$ are given by
\begin{equation}
    \ket{\nu_\alpha(t)} = 
    \sum_{i=1}^n U_{\alpha i}^* \int dp\ \psi_i(p) e^{-iE_i(p)t} \ket{\nu_i(p)}.
\end{equation} Here the produced state is a superposition of mass eigenstates with defined momenta, described by the wave function in momentum space, $\psi_i(p)$. This quantum state is localized in space and can correctly describe the physics of a propagating neutrino.
In order to reach a practical result, we assume the evolution to be one dimensional and the momentum distribution to be Gaussian.

After its propagation, the produced neutrino $\nu_\alpha$ can be detected at some detector in a position $L$ at time $T$ through a charged-current interaction $\nu_\alpha X\to l_\beta Y$, with $l_\beta$ a lepton of flavor $\beta$. The amplitude for this process in the WP formalism is \cite{Giunti:1997wq}\vspace{-0.1 cm}
\begin{equation}
    A_{\alpha\beta} \propto \sum_{i=1}^n U_{\alpha i}^* U_{\beta i}\exp\left\{-iE^0_i T + iP_iL - \frac{(L-v_iT)}{4\sigma_x^2} \right\}.
\end{equation}
Here $P_i$ is the central linear momentum of each mass eigenstate wave packet, $E_i^0 = \sqrt{P_i^2 + m_i^2}$ its central energy and $v_i = \partial E_i(p)/\partial p|_{p=P_i}$ its group velocity. Finally, $\sigma_x$ is a length scale which parametrizes the dampening of the oscillatins and that can be referred as the wave packet size~\cite{Giunti:1997wq, Nussinov:1976uw, Kayser:1981ye, Kiers:1995zj, Beuthe:2001rc,Akhmedov:2012uu,Akhmedov:2017mcc} This wave packet size depends on the neutrino production and detection mechanisms.

Most experiments do not measure $T$ precisely and oscillation periods are always much smaller than the operation time of the detector. Thus, the total probability $P_{\alpha\beta}(L)=\int_0^\infty dT|A_{\alpha\beta}|^2$ depends only on $L$,
\begin{align} \label{eq:app:full-prob}
P_{\alpha\beta} =& \sum_{i=1}^n |U_{\alpha i}|^2|U_{\beta i}|^2 + 2\text{Re} \sum_{j>i}U_{\alpha i}U_{\alpha j}^*U_{\beta i}^*U_{\beta j}\exp\left\{-2\pi i\frac{L}{\Losc^{ij}}-2\pi^2\left(\frac{\sigma_x}{\Losc^{ij}}\right)^2 - \left(\frac{L}{\Lcoh^{ij}}\right)^2\right\}.
\end{align} Here we have imposed \textit{a posteriori} the conservation of probability $\sum_\alpha P_{\alpha\beta}=1$ and have defined
\begin{equation} 
    \Losc^{ij} = \frac{4\pi E}{\Delta m^2_{ji}}\quad \text{and}\quad
    \Lcoh^{ij} = \frac{4\sqrt{2}E^2\sigma_x}{\Delta m^2_{ji}},
\end{equation}the oscillation and coherence lengths, respectively. This formula can be obtained in a more consistent manner in QFT formalism~\cite{Beuthe:2001rc,Giunti:2002xg,Akhmedov:2010ms,Torres:2020gzm}, without any \textit{a posteriori} conservation of probability (and an additional energy dependence).

Experiments based on nuclear decays only study the survival probability of electron antineutrinos $P(\nu_e\to\nu_e) \equiv P_{ee}$. Following from~\cref{eq:app:full-prob}, the full probability of this process is
\begin{align}\label{eq:app:surv-prob}
   P_{ee} = 1&- \sin^2 2\theta_{12}\cos^4\theta_{13}\cos^4\theta_{14} \Delta_{21} \\ &-
   \sin^22\theta_{13}\cos^4\theta_{13}(\cos^2\theta_{12}\Delta_{31}+\sin^2\theta_{12}\Delta_{32}) \nonumber \\ &- \sin^22\theta_{14}\big[\cos^2\theta_{13}\cos^2\theta_{12}\Delta_{41} +
   \cos^2\theta_{13}\sin^2\theta_{12}\Delta_{42} + \sin^2\theta_{13}\Delta_{43}\big] ,\nonumber
\end{align}where we have defined, similarly to ~\cite{Giunti:1997wq},
\begin{equation}
    \Delta_{ji} = \frac{1}{2}\left(1-\cos\frac{L\Delta m_{ji}^2}{2E}\exp\left\{-\frac{L^2(\Delta m^2_{ji})^2}{32E^4\sigma_x^2}\right\} \right).
\end{equation}We would like to emphasize that \cref{eq:app:surv-prob} is indeed the survival probability implemented in our analyses. On the contrary, \cref{eq:surv_prob} is only a concise approximation which is reasonably valid for short baseline experiments. In the PW formalism one obtains $\Delta_{ji} = \sin^2(L\Delta m^2_{ji}/4E)$,\ie, the same result that \cite{DayaBay:2016qvc,Andriamirado2021,Barinov:2021asz,NEOS:2016wee} use.

\section*{Details of the data analysis}
In the present work we have performed five different data analyses. Here we detail the main differences between them. 

\subsection{Daya Bay analysis}
\label{sec:dayabay}
For the analysis of the Daya Bay data~\cite{DayaBay:2016ggj} we have defined a test statistic based on a Poisson log-likelihood,
\begin{align}\label{eq:app:DBchi2}
    \mathcal{TS}^{\rm DayaBay}(\Delta m^2_{41},\theta_{14}, \vec\alpha)  = -2\sum_d\sum_{i=1}^{35}\biggr(&O^{d}_i-[\alpha_i N^{d}_i(\Delta m^2_{41},\theta_{14})+B^{d}_i] +  O^{d}_i\log\frac{\alpha_i N^{d}_i(\Delta m^2_{41},\theta_{14}) + B^{d}_i}{O^{d}_i}\biggr)\, .
\end{align}This statistic is defined from the Poisson probability $P(k,\lambda) = {e^{-\lambda}\lambda^k}/{k!}$ and already takes into account statistical uncertainties, which are dominant in the Daya Bay experiment. 
In~\cref{eq:app:DBchi2} $O^{d}_i,\, B^{d}_i,\, N^{d}_i$ are the observed, background, and predicted data in the energy bin $i$ and experimental hall $d= \text{EH1},\, \text{EH2},\, \text{EH3}$, respectively. 

The reactor flux in which the analysis is built is taken from the theoretical predictions of Huber and Mueller~\cite{Huber:2011wv,Mueller:2011nm}, even though there are known anomalies to them~\cite{Huber2016}. 
Then, $\vec\alpha$ are nuisance parameters that accommodate the uncertainties in this flux. These are different for each energy bin but the same for each experimental hall and minimize~\cref{eq:app:DBchi2},
\begin{equation}
\alpha_i = \frac{\sum_{d}O^{d}_i-B^{d}_i}{\sum_{d}N^{d}_i}\, .
\end{equation}With these nuisance parameters, the source flux and its normalization are free and the same for the three experimental halls. Only relative differences between detectors (\eg, neutrino oscillations) will be manifest. 

$O^{d},\, B^{d}$ are taken from the Supplemental Material of~\cite{DayaBay:2016ggj}, while $N_i^{d}$ is computed following~\cite{Dentler:2017tkw}
\begin{equation}\label{eq:app:DBexpected}
N_i^{d} = \mathcal{N}^{d} \sum_{\text{r}} \frac{\epsilon^d}{L_{r,d}^2}\int_{E^{\text{rec}}_{i}}^{E^{\text{rec}}_{i+1}}dE^{\text{rec}} \int_0^\infty dE_\nu\, \sigma(E_\nu)\,\phi(E_\nu)\, P_{ee}^{r,d}(E_\nu)\, R(E^{\text{rec}},E_\nu)\, .
\end{equation}Here, 
\begin{itemize}
    \item $\mathcal{N}^{d}$ is a normalization constant which takes into account the number of target protons in the detector. Note that this factor is accommodated in~\cref{eq:app:DBchi2} by the free nuisance parameters $\vec\alpha$ and therefore plays no role. However, we choose it such that our prediction of the expected events without oscillations match the corresponding data from Daya Bay.
    \item $r$ runs over the different reactor neutrino sources.
    \item $\epsilon^{d}$ is the detection efficiency of the experimental hall (averaged over all the detectors in the experimental hall), taken from Table VI in~\cite{DayaBay:2016ggj}.
    \item  $L_{r,d}$ is the mean distance between the reactor and the detectors in the experimental hall, taken from Table I in~\cite{DayaBay:2016ggj}.
    \item $E^{\text{rec}},\, E_\nu$ stand for the reconstructed and true neutrino energies.
    \item $\sigma(E_\nu)$ is the inverse beta decay cross section~\cite{Oralbaev_2016}.
    \item $\phi(E_\nu)$ is the Huber-Mueller flux~\cite{Huber:2011wv,Mueller:2011nm},
    \begin{equation}
        \phi(E_\nu) = \sum_{\text{isotope}} f_{\text{isotope}}\, \phi^{\text{isotope}}(E_\nu)\, ,
    \end{equation} 
    with $f_{\text{isotope}}$ the mean fission fraction of isotope $=\,  ^{235}\text{U},\, ^{238}\text{U},\, ^{239}\text{Pu},\, ^{241}\text{Pu}$.
    \item $P^{r,\text{exp}}_{ee}$ is the survival probability from~\cref{eq:app:full-prob}.
    \item $R(E^{\text{rec}},E_\nu)$ is the response matrix of the Daya Bay detectors~\cite{DayaBay:2016ggj}.
\end{itemize}
Note that in this analysis (and all the following) the standard oscillation parameters are not free but fixed at the values of~\cite{Esteban2020}. A more rigorous treatment would marginalize $\theta_{13}$ and $\Delta m_{31}^2$. However, the effect would be small even in the worst-case scenario and thus the present work does not consider this marginalization.

\subsection{NEOS analysis}
\label{sec:NEOS}
Our NEOS analysis is built from the data in Fig.~3(c)~\cite{NEOS:2016wee} and consists in the $\chi^2$ function
\begin{equation}\label{eq:app:NEOSchi2}
    \mathcal{TS}^{\rm NEOS}(\theta_{14},\Delta m^2_{41},\vec \alpha) = \sum_{i,j = 1}^{60}\left(R_i - \frac{N_i(\Delta m^2_{41},\theta_{14})+B_i}{N_i^{\text{SM}}+B_i}\right)(V^{-1})_{ij}\left(R_j - \frac{N_j(\Delta m^2_{41},\theta_{14})+B_j}{N_j^{\text{SM}}+B_j}\right)\, .
\end{equation}Here, $R_i$ is the ratio data from Fig.~3(c) in~\cite{NEOS:2016wee}, $B_i$ is the background events from Fig.~3(a), $V_{ij}$ is the NEOS covariance matrix, and $N_j(\Delta m^2_{41},\theta_{14}),\, N_j^{\text{SM}}$ are the expected events at NEOS with a 3+1 and a 3 neutrino model, respectively. Since there is only one detector, the nuisance parameters are fixed to unity.

The expected number of values are obtained using
\begin{equation}
N_i = \mathcal{N} \int_{L_{\text{min}}}^{L_{\text{max}}} \frac{dL}{L^2}\int_{E^{\text{rec}}_{i}}^{E^{\text{rec}}_{i+1}}dE^{\text{rec}} \int_0^\infty dE_\nu\, \sigma(E_\nu)\,\phi^{\text{DB}}(E_\nu)\, P_{ee}(L,E_\nu)\, R(E^{\text{rec}},E_\nu)\, .
\end{equation}Here, 
\begin{itemize}
    \item The normalization factor $\mathcal{N}$ is free and adjusted to match the total number of observed events from Fig.~3(a) for any $(\Delta m^2_{41},\theta_{14})$, taking into account the background.
    \item Since the baseline is short, finite-size effects of the detector need to be taken into account by integrating between $L_{\text{min}} = 22.14$ m and $L_{\text{max}} = 25.14$ m.
    \item $\sigma(E_\nu)\phi^{DB}(E_\nu)$ is the Daya Bay antineutrino flux weighted by the inverse beta decay cross section, taken from Table 12 in~\cite{DayaBay:2018heb}. As noted in~\cite{Dentler:2017tkw}, this spectrum is computed under the assumption of three-flavor oscillations, and thus these oscillations, although small, should be unfolded for a rigorous analysis. This effect is only corrected in the Daya Bay + NEOS analysis.
    \item The response matrix $R(E^{\text{rec}},E_\nu)$ is not provided by the NEOS collaboration, and therefore has to be reproduced from~\cite{NEOS:2017,RENO:2021} using the same technique as in~\cite{Huber2016}.
\end{itemize}

Finally, to build $V_{ij}$ we have digitized the correlation matrix from \cite{NEOS:2017}, which has unity diagonal elements. Then, this matrix is rescaled such that its diagonal elements match the quadratic sum of the systematical and statistical errors digitized from Fig.~3(c) in \cite{NEOS:2016wee}. We take this rescaled matrix to be the covariance matrix $V_{ij}$ in (\ref{eq:app:NEOSchi2}).

\subsection{Daya Bay + NEOS analysis}
\label{sec:dayabayNEOS}
Now NEOS is treated as if it was a fourth Daya Bay detector. That is, we have computed the expected events using the same Huber-Mueller flux for both Daya Bay and NEOS, and accommodated the flux uncertainties using a common vector of nuisance parameters. 

However, we must take into account that the energy bins for Daya Bay and NEOS are different. On the one hand, Daya Bays energy range is $E^\text{rec}_{\text{DB}}\in (0.7,12.0){\rm MeV}$, with energy binning $\Delta E^{\text{rec}}_{\text{DB}} = 0.2$ MeV. On the other hand, NEOS measures in $E^\text{rec}_{\text{NEOS}}\in (1.0,10.0){\rm MeV}$ and with bins of width $\Delta E^{\text{rec}}_{\text{NEOS}} = 0.1$ MeV. Therefore, we pick the conservative choice to only consider the energy bins that are well defined in both experiments and that share the same energy bin edges, \ie, $E^\text{rec}_{\text{DB}}\in (1.3,6.9){\rm MeV}$. Then, NEOS has twice as many bins as Daya Bay. This also affects the definition of the nuisance parameters: the nuisance parameter $\alpha_i$ of Daya Bay's energy bin $i$ is applied to two consecutive energy bins in NEOS.

Taking all this into account, the test statistic to minimize is
\begin{align}
     \mathcal{TS}^{\rm DayaBay+NEOS}=&   -2\sum_{d}\sum_{i=2}^{29}\biggr(O^{d}_i-[\alpha_i N^{d}_i(\Delta m^2_{41},\theta_{14})+B^{d}_i] + O^{d}_i\log\frac{\alpha_i N^{d}_i(\Delta m^2_{41},\theta_{14}) + B^{d}_i}{O^{d}_i}\biggr) + \\ \nonumber & +
     \sum_{i,j = 4}^{59}\left(R_i - \frac{\alpha_{\text{floor}(i/2)}N_i(\Delta m^2_{41},\theta_{14})+B_i}{N_i^{\text{SM}}+B_i}\right) (V^{-1})_{ij}\left(R_j - \frac{\alpha_{\text{floor}(j/2)}N_j(\Delta m^2_{41},\theta_{14})+B_j}{N_j^{\text{SM}}+B_j}\right)
     \, ,
\end{align}where $\mathcal{TS}^{\rm DayaBay+NEOS} \equiv  \mathcal{TS}^{\rm NEOS}(\theta_{14},\Delta m^2_{41},\vec \alpha) +\mathcal{TS}^{\rm DayaBay}(\theta_{14},\Delta m^2_{41},\vec \alpha)$ is the test statistic presented in~\cref{eq:test-statistic}. Now, the minimization of $\vec\alpha$ can only be done numerically.

\subsection{PROSPECT analysis}
\label{sec:PROSPECT}
The analysis of the PROSPECT data~\cite{Andriamirado2021} is independent from those of Daya Bay and NEOS, since PROSPECT's neutrino source only contains $^{235}$U. Thus, it is not reasonable to use the same source flux for all experiments. The PROSPECT detector is subdivided onto independent segments at difference distances to the nuclear reactor. These segments are capable of measuring neutrino propagation in different baselines, and are sensitive to a $\SI{1}\eV^2$ sterile neutrino oscillation.

The test statistic to minimize is the $\chi^2$ function
\begin{equation}\label{eq:app:PROSPECTchi2}
\mathcal{TS}^{\rm PROSPECT}(\theta_{14},\Delta m^2_{41}) = \Vec{x}\cdot V^{-1}\cdot\Vec{x}\, ,
\end{equation}where $V$ is the PROSPECT covariance matrix, and $\Vec{x}$ is a 160-dimensional vector that describes the discrepancy between data and prediction. Namely, it contains this information at each of the 16 energy bins of each of the 10 different baselines, ordered in increasing length, and then in increasing energy. For each baseline $l$ and energy bin $e$, it is defined as
\begin{equation}
    x^{l,e} = M^{l,e} - M^e \frac{P^{l,e}}{P^e}\, . 
\end{equation}Here, $M^{l,e}, \, P^{l,e}$ are the observed and the predicted data at baseline $l$ and energy bin $e$, respectively. Then, $M^e,\, P^e$ represent the total observed and predicted data, respectively, summing for all baselines. That is,
\begin{equation}
    M^e = \sum_{l=1}^{10}M^{l,e}\quad \text{ and } \quad
    P^e = \sum_{l=1}^{10}P^{l,e}\, .
\end{equation}

The test statistic in~\cref{eq:app:PROSPECTchi2} minimizes the effect of source flux uncertainties and is independent of its normalization. Therefore, we use the Hubber-Mueller flux for $^{235}$U only, $\phi_{\text{235U}}$~\cite{Huber:2011wv}. The prediction is computed as
\begin{equation}
P^{l,e} = \mathcal{N}\sum_{\text{seg}\in l} \epsilon^{\text{seg}} \int_{L_{\text{seg}}-\delta L}^{L_{\text{seg}}+\delta L}\frac{dL_{\text{seg}}}{L_{\text{seg}}^2}\int_0^\infty dE_\nu \sigma(E_\nu)\phi_{\text{235U}}(E_\nu) P_{ee}(L_{\text{seg}},E_\nu)R(E^e,E_\nu)\, ,
\end{equation}where $E^e$ is the central energy of the energy bin $e$, $L_{\text{seg}}$ is the baseline of the segment, $\epsilon^{\text{seg}}$ its efficiency, $R$ the response matrix provided by the collaboration, and the sum is done for all segments in the same baseline~\cite{Andriamirado2021}. We also perform a fast integration in $L_{\text{seg}}$ to consider the finite width of the reactor and the segments, with $\delta L = 0.25{\rm cm}$. Although the normalization constant $\mathcal{N}$ plays no role in~\cref{eq:app:PROSPECTchi2}, it is computed such that our predicted data without oscillations matches the analogous PROSPECT results at each baseline.

\subsection{BEST analysis}
\label{sec:BEST}
Again, the analysis on the Baksan Experiment on Sterile Transition data~\cite{Barinov:2021asz} is independent from the rest of experiments. BEST uses a $^{51}$Cr radioactive source, which emits neutrinos in only four discrete energies, namely $E_i = 747,\, 427,\, 752,\, 432\,\mathrm{keV}$. Their fission fractions are $f_i = 0.8163, 0.0895, 0.0849, 0.0093$, respectively. 

Our $\chi^2$ only takes into accounts two points, namely
\begin{equation}\label{eq:app:BESTchi2}
\chi^2_{\text{BEST}}(\Delta m^2_{41},\theta_{14}) = \frac{(r^{\text{in}}_{\text{meas}}-r^{\text{in}}_{\text{pred}})^2}{\epsilon_{\text{in}}^2} + \frac{(r^{\text{out}}_{\text{meas}}-r^{\text{out}}_{\text{pred}})^2}{\epsilon_{\text{out}}^2}\, .
\end{equation}Here, $\epsilon$ are the statistical and systematic uncertainties, and $r$ are the measured and predicted production mean rates. The predicted rate $r_{\text{pred}}$ is computed as 
\begin{equation}\label{eq:app:production-rate}
    r_{\text{pred}} = \xi_{\text{in/out}}\frac{n\sigma A_0}{4\pi}\int_{V_{\text{in/out}}} \frac{\sum f_i P_{ee}(L,E_i)}{L^2}dV\, ,
\end{equation}with $n = (2.1001\pm 0.0008)\times 10^{22}/\text{cm}^3$ the $^{71}$Ga number density of the detector, $\sigma = (5.81^{+0.21}_{-0.16})\times 10^{-45}\text{ cm}^2$~\cite{Barinov:2021asz,Bahcall1997} the neutrino capture cross section, $A_0 = (3.414\pm 0.008)\text{ Ci}$ the initial activity of the $^{51}$Cr source, and the integration is done for the whole volume of the inner or the outer detector. The geometry of the inner and outer detectors are not exactly known and are subject to experimental details such as the quantity of $^{71}$Ga or the position of tubes inside the detector. Therefore, we add two geometric correction factors $\xi_{\text{in/out}}$. The BEST data provide the values of the integrals in~\cref{eq:app:production-rate} when $P_{ee} = 1$. We pick $\xi_{\text{in/out}}$ to match these values, and neglect its dependence on $P_{ee}$.  

In~\cref{eq:app:BESTchi2} the production rate predictions are compared with the $r_{\text{meas}}$ from Table I in~\cite{Barinov:2021asz}. Namely, $r_{\text{meas}}^{\text{in}}=54.9^{+2.5}_{-2.4}$ and $r_{\text{meas}}^{\text{out}} = 55.6^{+2.7}_{-2.6}$. Finally, $\epsilon^2$ are computed as the square sum of statistical uncertainties (taken from Table I \cite{Barinov:2021asz}), systematic uncertainties ($\sim 2$\%) and the cross section uncertainty.
\end{document}